\documentclass[10pt,twocolumn,showpacs,preprintnumbers,amsmath,amssymb,%
aps,prd,a4paper,floatfix,nofootinbib,showkeys]{revtex4-1}  

\usepackage{amssymb,amsfonts,amsmath,bbm,dsfont}
\usepackage{graphicx}
\usepackage{hyperref}
\usepackage{multirow}
\usepackage{color}

\usepackage[dvipsnames]{xcolor}

\makeatletter

\renewcommand{\@makefntext}[1]{\parindent=1em\noindent\hbox to 1.8em{\hss$^{\@thefnmark}$}#1}
\renewcommand{\@footnotemark}{\hbox{\mathsurround=0pt$^{\@thefnmark}$}}

\makeatother

\usepackage[tight]{units}

\newcommand{\absvec}[1]{\left| \vec{#1} \right|}

\newcommand{\be}{\begin{equation}}
\newcommand{\ee}{\end{equation}}

\renewcommand{\vec}[1]{\boldsymbol{#1}}

\newcommand{\tr}[1]{\,\text{tr}\left[#1\right]}

\newcommand{\identity}{\mathrm{I}}
\newcommand{\sign}{\mathrm{sign}}

\begin{document}
\date{\today}

\title{Coulomb vs. physical string tension on the lattice}
\author{Giuseppe~Burgio}
\affiliation{Institut f\"ur Theoretische Physik, Auf der Morgenstelle 14,
 72076 T\"ubingen, Germany}

\author{Markus~Quandt}
\affiliation{Institut f\"ur Theoretische Physik, Auf der Morgenstelle 14, 
72076 T\"ubingen, Germany}
\author{Hugo~Reinhardt}
\affiliation{Institut f\"ur Theoretische Physik, Auf der Morgenstelle 14, 
72076 T\"ubingen, Germany}

\author{Hannes~Vogt}
\affiliation{Institut f\"ur Theoretische Physik, Auf der Morgenstelle 14, 
72076 T\"ubingen, Germany}

\begin{abstract}
From continuum studies it is known that the Coulomb string tension $\sigma_C$
gives an 
upper bound for the physical (Wilson) string tension $\sigma_W$ 
\cite{Zwanziger:2002sh}. 
How does however such relationship translate to the lattice?
In this paper we give evidence that there, while the two string tensions 
are related at zero temperature, they decouple at finite temperature. 
More precisely, we show that on the lattice the Coulomb gauge 
confinement scenario is always tied to the {\it spatial} string tension, which 
is known to survive the deconfinement phase transition 
and to cause screening effects in the quark-gluon plasma. Our analysis is 
based on the identification and elimination of \emph{center vortices} which
allows to control the physical string tension and study its effect on the 
Coulomb gauge observables. We also show how alternative definitions of the 
Coulomb potential may sense the deconfinement transition; however a true
static Coulomb gauge order parameter for the phase transition is still elusive 
on the lattice.
\end{abstract}
\maketitle

\section{Introduction}

Recent years have seen a rising interest in Coulomb gauge 
investigations of Yang-Mills theories in general,
and in the Hamiltonian formulation in particular
\cite{Jackiw:1977ng,Schutte:1985sd,Szczepaniak:2001rg,Feuchter:2004mk,%
Schleifenbaum:2006bq,Epple:2006hv,Epple:2007ut,Feuchter:2007mq,%
Reinhardt:2008ij,Reinhardt:2008ek,Campagnari:2010wc,Pak:2011wu,%
Campagnari:2011bk,Reinhardt:2011hq,Heffner:2012sx,Reinhardt:2012qe,%
Reinhardt:2013iia}.
In the latter, once Weyl-gauge is implemented to eliminate the $A_0(x)$ 
components of the gauge fields, the Hamilton operator and the Gau\ss's law 
constraint are invariant under the residual time-independent gauge 
transformations and, moreover, only depend on the remaining space-like 
gauge fields and momenta $\vec{A}^a(\vec{x})$, $\hat{\vec{\Pi}}^a(\vec{x})$.
In Abelian theories the \emph{transversal} part of these vector fields 
is gauge-independent and thus physical, so that \emph{Coulomb gauge} can
be seen as \emph{the} physical gauge, eliminating all gauge-dependent
degrees of freedom. In non-Abelian theories this is no longer strictly 
true, but the elimination of the longitudinal degrees of freedom via
Coulomb gauge still 
resolves Gau\ss{}' law, providing a formulation in terms of the 
transversal fields\footnote{We will omit the index $\perp$ on transversal
vector fields in the following.} $\vec{A}_\perp^a$, $\vec{\Pi}^a_\perp$ alone,
where studies of the Yang-Mills ground state are more natural.
Such a resolution of Gau\ss{}' law through Coulomb gauge thus ``automatically'' 
incorporates the constraints one should impose in the Hamiltonian
formulation,
circumventing the explicit construction of the physical 
Hilbert space \cite{Burgio:1999tg}. This results in:
\begin{align}
H &= H_{\textsc{G}} + H_{\textsc{C}} \; , \\
\label{G-Ham}
H_{\textsc{G}} &= \int d^3 x  \left[ \frac{1}{2} \mathcal{J}^{-1}[\vec{A}]
\,\Pi^{a}_i \, \mathcal{J}[\vec{A}]\,\Pi^{a}_i + 
\frac{1}{4} F^a_{i j} \,F^a_{i j} \right] \; , \\
\label{Coulomb-Ham} 
H_{\textsc{C}} &= \frac{g^2}{2} \int d^3(x,y)\,
\mathcal{J}^{-1}[\vec{A}]\, \rho^a(\vec{x})  \,\mathcal{J}[\vec{A}]
\hat{F}^{a b}(\vec{x},\vec{y}) \rho^b(\vec{y}) \; ,
\end{align}
where $\mathcal{J}[\vec{A}]$ is the determinant of the 
Faddeev-Popov operator, i.e. the inverse Coulomb ghost propagator
\begin{align}
\label{G-inv}
(\hat{G}^{-1})^{a b}(\vec{x},\vec{y}) = (-\partial_i \hat{D}_i^{a b} ) \delta(\vec{x}-\vec{y}) \;,
\end{align}
while the Coulomb Hamiltonian $H_{\textsc{C}}$ 
describes
the self-interaction of non-Abelian color charges with density
\begin{align}
\label{rho}
 \rho^a(\vec{x}) = \psi^{\dagger}(\vec{x}) T^a \psi(\vec{x}) - f^{a b c} 
A_i^b(\vec{x}) \Pi_i^c(\vec{x})
\end{align}
through the non-Abelian Coulomb kernel
\begin{align}
\label{Coulomb-kernel-0}
 \hat{F}^{a b}(\vec{x},\vec{y}) =  \int d^3 z \, G^{a c}(\vec{x},
\vec{z}) (- \partial_{\vec{z}}^2) G^{c b}(\vec{z},\vec{y})  \; .
\end{align}
The first term on the right-hand side of Eq.~\eqref{rho} is the matter
charge density, which for the pure Yang-Mills case should
be understood as an external source, while the second part is the 
dynamical charge density of the non-Abelian gauge field. 
In the Abelian theory the latter would, of course, be absent and 
Eq.~\eqref{Coulomb-kernel-0} becomes the ordinary Coulomb kernel, i.e. 
the Green's function of the Laplacian 
$\hat{F}(\vec{x},\vec{y}) =(4\pi|\vec{x}-\vec{y}|)^{-1}$.

From Eq.~\eqref{Coulomb-kernel-0} one can
define the non-Abelian color Coulomb potential, i.e.~the Coulomb energy 
density for a pair of static quark-antiquark color charges 
separated by a distance $\vec{x}$:
\begin{align}
\label{V-prop}
V_{\textsc{C}}^{a b}(\vec{p}) =  g^2 \, \int d^3 x \, e^{-i \vec{p} \cdot
 \vec{x}} \langle \hat{F}^{a b}(\vec{x},\boldsymbol{0}) \rangle \;.
\end{align}
In a seminal paper \cite{Zwanziger:2002sh} Zwanziger, extending
ideas first put forward by Gribov \cite{Gribov:1977wm}, showed how such 
a Coulomb potential gives a natural upper bound to Wilson's physical
potential \cite{Wilson:1974sk}.
In other words, the presence of Coulomb confinement is a {\it necessary} 
condition for the physical confinement mechanism to take place 
in Yang-Mills theories. These results are based on Gribov's 
intuition that the Yang-Mills dynamics must be restricted to the first 
Gribov region, where the Faddeev-Popov operator in Eq.~\eqref{G-inv}
is positive definite.\footnote{A unique elimination of all gauge copies 
requires an even further restriction to the so-called fundamental modular 
region, where the gauge functional only possesses absolute maxima.}
Further signatures of this idea are the infra-red (IR) divergence of the 
Coulomb gauge ghost form factor and the emergence of an IR scale 
in the gluon dispersion relation \cite{Gribov:1977wm,Cornwall:1979hz}.

The Gribov-Zwanziger confinement scenario has been investigated 
in detail on the lattice
\cite{Zwanziger:1993dh,Zwanziger:1995cv,Cucchieri:2000gu,Cucchieri:2000kw,%
Greensite:2003xf,Langfeld:2004qs,Greensite:2004ke,Burgio:2008jr,Voigt:2008rr,%
Greensite:2009eb,Burgio:2009xp,Nakagawa:2008zza,Nakagawa:2009is,Quandt:2010yq,%
Nakagawa:2010eh,Iritani:2010mu,Reinhardt:2011fq,Nakagawa:2011ar,Burgio:2012bk,%
Burgio:2012ph,Iritani:2012bc}, confirming the expected relationships between 
Coulomb gauge Greens-functions, Coulomb potential and confinement. 
However, since all lattice investigations 
are defined through an Euclidean-Lagrangian formalism, the 
contact with results obtained in a continuum Hamiltonian formulation is 
not straightforward. In particular, Weyl gauge can {\it never} be implemented
on the lattice due to the periodic boundary conditions in the time direction,
even in the so-called lattice Hamiltonian limit 
\cite{Burgio:2003in,Burgio:2012bk}, where one takes
strongly anisotropic lattices with a much finer spacing
in the temporal direction. At finite temperatures, due to the 
fixed finite length of the compactified time direction, the situation
gets even worse.

The main problem is that in any Euclidean-Lagrangian formalism static 
quantities must be extracted from correlators which extend along the time 
direction.\footnote{For instance, Polyakov loops are a most efficient way to determine 
the static inter-quark 
potential \cite{Luscher:2001up}.} Lattice Coulomb gauge observables, on the 
other hand, are defined at fixed time slices, involving only the space 
components of the vector fields. 
At $T=0$ the $O(4)$ rotational symmetry is unbroken and the restriction to 
space-like gauge fields is irrelevant, so that the Coulomb gauge analysis  of 
confinement 
on the lattice is fully valid. As $T$ increases, however, it is conceivable 
that lattice Coulomb gauge observables, remaining ``stuck'' into the fixed 
time slice, will only 
sense space-space correlations. Since the area-law for {\it spatial} Wilson 
loops 
survives above $T_c$, causing screening effects in the quark gluon plasma, 
this non-perturbative effect could turn out to dominate the Coulomb gauge 
dynamics 
well above the deconfinement transition. In fact, all attempts 
to extend lattice investigations in 
Coulomb gauge to finite temperature \cite{Vogt:2013jha} 
have up to now led to inconclusive results. 

In this paper we give evidence that:
\begin{itemize}
\item on the lattice, the relationship between Gribov-Zwanziger and 
Wilson confinement disappears above the deconfinement phase transition;
\item the reason for such a failure lies in the
strong correlation between the Coulomb string tension $\sigma_C$ 
and the \emph{spatial} Wilson string tension.
\end{itemize}
Moreover, to calculate $\sigma_C$ one also needs to discretize the Coulomb 
potential. Contray to the continuum Hamiltonian formulation, on a finite 
lattice several \emph{inequivalent} definitions of $V_C$ are possible. 
Our numerical results lead to the conclusion that extending the 
meaning of the Coulomb potential $V_C$ as the force between colour charges
from the Hamiltonian picture to the lattice formulation can lead to 
inconsistent results and that the lattice versions of $V_C$ are sensitive to 
the same quark correlations that build the spatial string tension in the high 
temperature phase

To test our assumption, we need {a tool} to control the Wilson string tension
$\sigma_W$, 
both for the whole ensemble and for its spatial directions separately. To do 
so, we adapt a method pioneered in Ref.~\cite{deForcrand:1999ms} by either 
removing:
\begin{itemize} 
\item all center vortices from the gauge field 
\\
(\emph{full vortex removal});
\item only vortices that pierce space-like Wilson loops 
\\
(\emph{spatial vortex removal}).
\end{itemize}
The rationale behind this strategy is clear: physical confinement should be 
caused by percolating center vortices piercing time-like Wilson loops
\cite{DelDebbio:1996mh,Langfeld:1997jx,DelDebbio:1998uu,Engelhardt:1999fd,%
Greensite:2003xf,Quandt:2010yq}.\footnote{A precise relationship between such 
a gauge-fixed, so called P-vortices and the topological center vortices 
originally introduced by 't Hooft \cite{'tHooft:1979uj} is still missing. 
The interested reader is referred to Refs.~\cite{Hart:2000en,%
Kovacs:2000sy,deForcrand:2000fi,Barresi:2001dt,deForcrand:2001nd,%
Barresi:2002un,Reinhardt:2002mb,Barresi:2003jq,Barresi:2004qa,Barresi:2006gq,%
Burgio:2006dc,Burgio:2006xj,Burgio:2014yna} for further discussions on the 
subject.} Removing all center vortices will thus disable confinement in the 
Yang-Mills ensemble, while
removing spatial vortices only should therefore not affect the inter-quark
potential; in fact, Polyakov loops correlators only involve temporal links and 
thus remain \emph{exactly} unaffected by such a procedure. Any effect of 
spatial vortex removal on Coulomb gauge observables thus cannot be related 
to confinement and must, instead, be attributed to the disappearance of quark 
screening effects through the removal of the
spatial string tension. This would then be a direct proof that such an 
observables 
predominantly see the spatial correlations in the gluon plasma rather than 
the confining force between static colour charges.

\section{Setup}

\subsection{Lattice setup}
For our  Coulomb gauge investigation we will employ the 
anisotropic Wilson action \cite{Klassen:1998ua,Burgio:2003in} 
for the colour group $SU(2)$ as proposed in 
Refs.~\cite{Nakagawa:2009is,Nakagawa:2011ar,Burgio:2012bk}:
\begin{eqnarray}
S &=& \sum_x \biggl\{\beta_s \sum_{j> i=1}^3\left(1-
\frac{1}{2} \mathrm{Re}\tr{U_{ij}(x)}\right)\nonumber\\
&+&\beta_t 
\sum_{i=1}^3
\left(1-\frac{1}{2}\mathrm{Re}\tr{U_{i4}(x)}\right)
\biggr\}\,,
\label{s_anis}
\end{eqnarray}
where $U_{\mu\nu}(x)$ is the standard plaquette. 
For each choice of $\beta_s \neq \beta_t$ the spatial and temporal
lattice spacings $a_s$ and $a_t$ have to be determined non-perturbatively, 
giving the renormalized anisotropy through the ratio $\xi = a_s/a_t$. 
The couplings are usually parameterized as $\beta_s = \beta \, \gamma$ and 
$\beta_t = \beta / \gamma$, where $\gamma$ is the bare 
anisotropy, which needs to be tuned with $\beta$ in order to realize the 
desired $\xi$ \cite{Burgio:2003in}. Tables for $\xi$ and 
$a_s$ at selected choices of $\beta$ and $\gamma$ can be found,
for the colour group $SU(2)$, in Ref.~\cite{Burgio:2012bk}.
All simulations for which no explicit value for $\xi$ is reported
have been performed in the isotropic case $\xi = 1$.

Most of the finite temperature simulations have been performed on lattices 
of sizes $V = N_t\times 32^3$ with varying $N_t$; different choices 
will be explicitly indicated in the data.
The gluon propagator, the ghost propagator and the Coulomb potential have been 
computed from 100 independent samples in double precision, while the precise 
determination of the string tension through Creutz ratios at large distances
required up to $10^5$ samples.

\subsection{Center vortex removal}
To identify center vortices, we first fixed the MC-configurations to 
the direct maximal center gauge \cite{DelDebbio:1998uu}, i.e.~we maximized
\begin{equation}
\label{eq:dmcg_functional}
 F[U] = \sum_{x,\mu}\tr{ U_\mu(x)^2}
\end{equation}
where $\mu = 0,1,2,3$ for the full (standard) maximal center gauge and 
$\mu = 1,2,3$ for the maximal center gauge restricted to the space-like
links (``spatial maximal center gauge''). 
For the numerical implementation of Eq.~\eqref{eq:dmcg_functional} 
we have used an iterated over-relaxation algorithm based on the gauge-fixing 
CUDA code cuLGT \cite{Schrock:2012fj}. For configurations which required 
subsequent Coulomb gauge fixing we stopped the center gauge fixing 
as soon as the functional value Eq.~\eqref{eq:dmcg_functional} changed by 
less than $\epsilon = 10^{-12}$ within 100 iterations. For the measurements
where no further gauge fixing was required we performed the center gauge 
fixing in single precision using $\epsilon = 10^{-7}$. 

Center \emph{projected} configurations are then obtained after 
center gauge-fixing by mapping the links to the closest center element:
\begin{equation}
 Z^\text{s/f}_\mu(x) = \sign\tr{U_\mu(x)}\, \identity\,,
\end{equation}
where the index ``s'' and ``f'' stands for ``spatial'' and ``full'', 
respectively,
with the index $\mu = 1, 2, 3$ in the former and $\mu = 0,\ldots,3$ 
in the latter case. To create vortex free configurations, 
we follow Ref.~\cite{deForcrand:1999ms} and define
\begin{equation}
 V^\text{s/f}_\mu(x) = Z^\text{s/f}_\mu(x) \cdot U_\mu(x)\,,
\end{equation}
where $\mu$ runs again over only spatial or all Lorentz indices, respectively. 

\subsection{Coulomb gauge}
Since we want to investigate the effect of vortex removal and center 
projection on correlators in Coulomb gauge, we need to transform each of the 
configurations $\{Z^\text{f}, Z^\text{s}, V^\text{f}, V^\text{s}\}$ 
discussed above to Coulomb gauge.\footnote{We are aware that this procedure 
is not exactly equivalent to fixing Coulomb gauge directly. For a critical 
discussions 
about the effect of subsequent incomplete gauge fixings, Gribov copies and 
projections 
see e.g.~Refs.~\cite{Bali:1996dm,Bornyakov:2000ig,Giusti:2001xf,Faber:2001hq,
Faber:2001zs,Boyko:2006ic,Cais:2008za}.} We employ a combination of simulated 
annealing and over-relaxation \cite{Bogolubsky:2005wf,Bogolubsky:2007bw}, 
again adapting the CUDA code cuLGT \cite{Schrock:2012fj}. 
For the center projected configurations we first had to apply a random gauge 
transformation, since the Coulomb FP-operator would otherwise be singular;
the links in the center-projected, Coulomb gauge-fixed configurations are 
therefore no longer elements of $\mathbbm{Z}_2$, but again of $SU(2)$.

After gauge-fixing we calculated, from the ghost propagator 
\begin{equation}
 G(\vec{p}) = \frac{d(\vec{p})}{\absvec{p}^2} = 
\mathrm{tr}\left\langle \left( -\hat{\vec{D}} \cdot \nabla \right)^{-1}
\right\rangle\,,
\end{equation}
the ghost form factor $d(\vec{p})$ and the Coulomb potential
\begin{equation}
\label{eq:coulombpotential}
V_C(\vec{p}) = g^2 \mathrm{tr}\left\langle \left( -\hat{\vec{D}} \cdot \nabla 
\right)^{-1} \left(-\nabla^2\right)\left( -\hat{\vec{D}} \cdot \nabla \right)^{-1}
\right\rangle\,,
\end{equation}
both directly in momentum space, where 
$\left( -\hat{\vec{D}} \cdot \nabla \right)$ is the Faddeev-Popov operator. 
If the Coulomb potential is linearly rising at large distances like 
$V_C(r) \simeq \sigma_C\,r$, its Fourier transform will behave as 
$V_C(p) \simeq 8 \pi \sigma_C / p^4$
at very small momenta. It is therefore customary to plot the quantity 
$p^4\,V_C(p)$ (often normalized by $8 \pi \sigma_W$) in which a non-zero 
intercept at 
$p \to 0$ signals a non-vanishing Coulomb string tension; we will follow this 
convention below. 

As mentioned in the introduction, there is also an alternative definition of 
the Coulomb potential, directly calculated in position space:
\begin{align}
\label{eq:correlator}
 aV_C(|\vec{x}-\vec{y}|) &= -\lim_{t\rightarrow 0} \frac{d}{dt} \log \left\langle 
\tr{P_t(\vec{x})P_t^\dagger(\vec{y})}\right\rangle\nonumber\\
  &= -\log \left\langle \tr{U_0(\vec{x})U_0^\dag(\vec{y})}\right\rangle\,,
\end{align}
where  $P_t(\vec{x})$ is a Polyakov line of length $(at)$ starting at lattice 
site $(0,\vec{x})$. The equality in the second line is not obvious and 
discussed in more detail in 
Ref.~\cite{Marinari:1992kh,Greensite:2003xf,Iritani:2010mu}.
Though formally equivalent in the Hamiltonian limit, such two alternative 
definitions of the Coulomb potential need not coincide on a finite lattice 
and will, in fact, show a rather different behaviour at finite temperatures. 
This is obviously due to the fact  that Eq.~\eqref{eq:coulombpotential} depends 
only on spatial links, whereas Eq.~\eqref{eq:correlator} depends only on 
temporal links; both get treated differently in Coulomb gauge {\it and}
at finite temperature.

\section{Results}

\subsection{Finite temperature in Coulomb gauge}
\label{sec:fT}
\begin{figure*}[htb]
	\center
	\includegraphics[width=0.99\columnwidth]{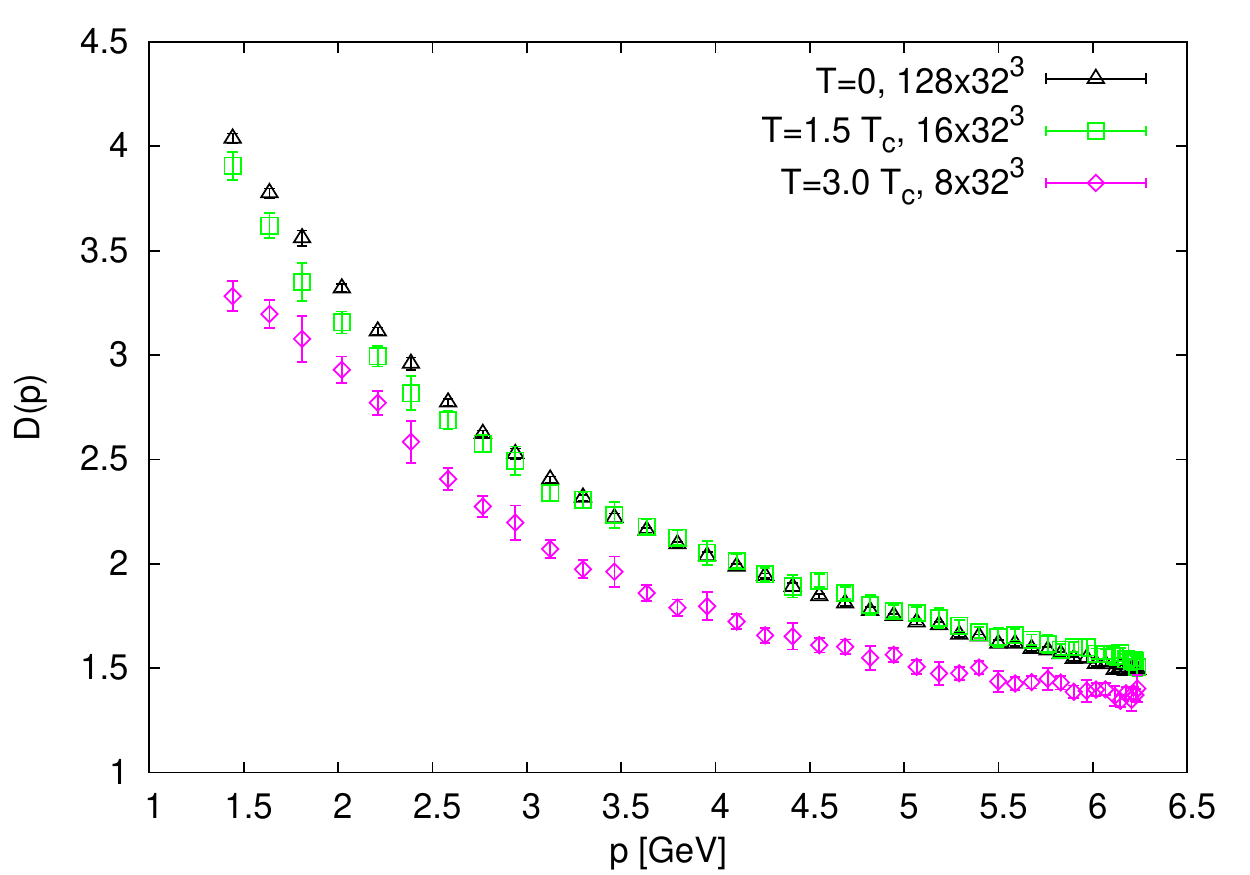}
	\caption{Gluon propagator for various temperatures at $\beta = 2.49$ 
                 and anisotropy $\xi=4$. As $T$ increases, the infrared 
                 Gribov mass $M \sim D^{-\frac{1}{2}}(0)$ increases as well
                 \cite{Burgio:2008jr,Burgio:2009xp}.}
	\label{fig1}
\end{figure*}

\begin{figure*}[htb]
	\center
	\includegraphics[width=0.99\columnwidth]{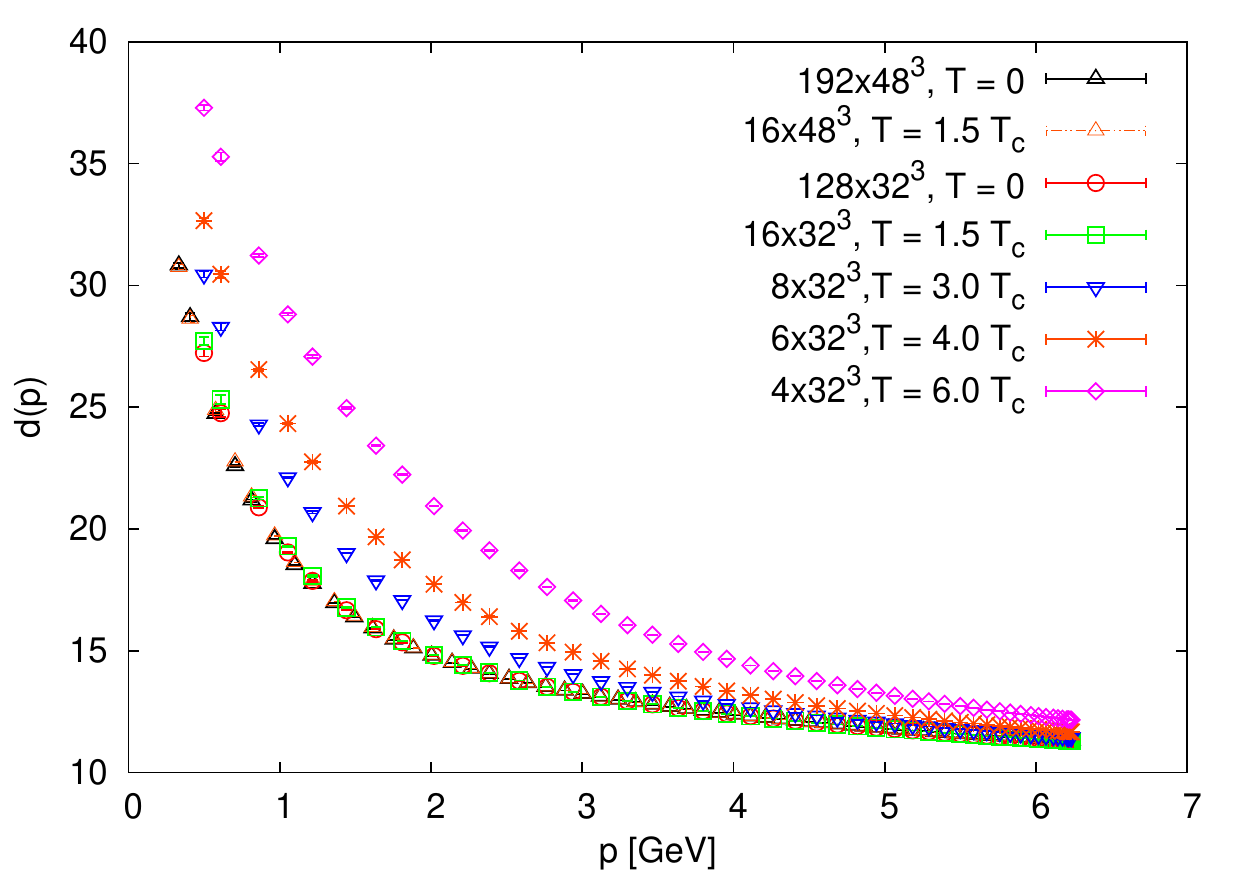}
	\caption{Ghost form factor for various temperatures at $\beta = 2.49$ 
        and anisotropy $\xi=4$. No flattening is observed as $T$ is increased,
        except perhaps for the $T= 6\, T_c$ data {(the upper curve), which 
        displays a slight decrease in the exponent of the infrared increasing 
        power law as compared to the other data}.}
	\label{fig2}
\end{figure*}

\begin{figure*}[htb]
	\center
	\includegraphics[width=0.99\columnwidth]{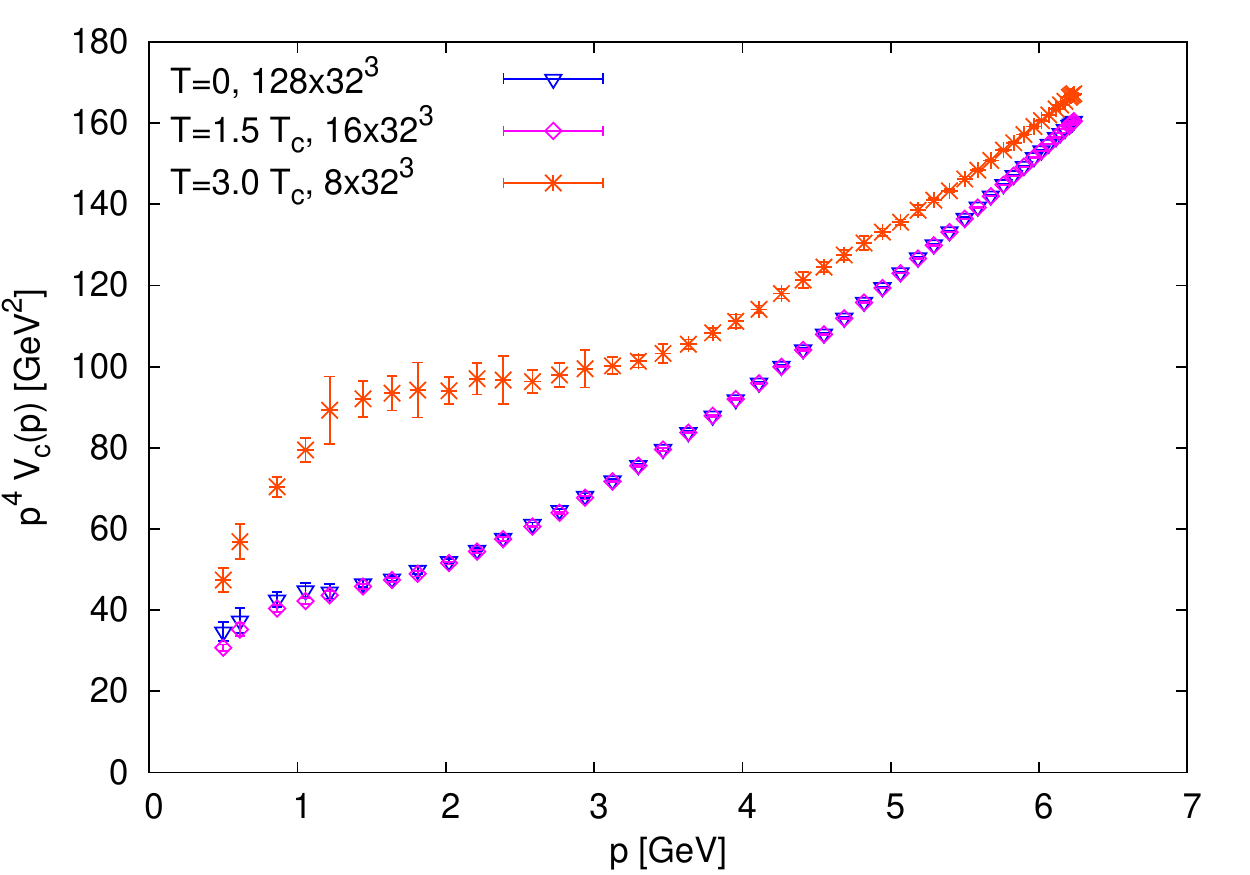}
	\caption{Coulomb potential from Eq.~(\ref{eq:coulombpotential}) 
	for various temperatures at $\beta = 2.49$ 
        and anisotropy $\xi=4$; {the normalization is such that the 
        intercept at $p \to 0$ equals $8 \pi \sigma_C$. We find that the}
        extrapolation to $p\to 0$ is compatible with an increase of 
        $\sigma_C$ at larger temperatures.}
	\label{fig3}
\end{figure*}

\begin{figure*}[htb]
	\center
	\includegraphics[width=0.99\columnwidth]{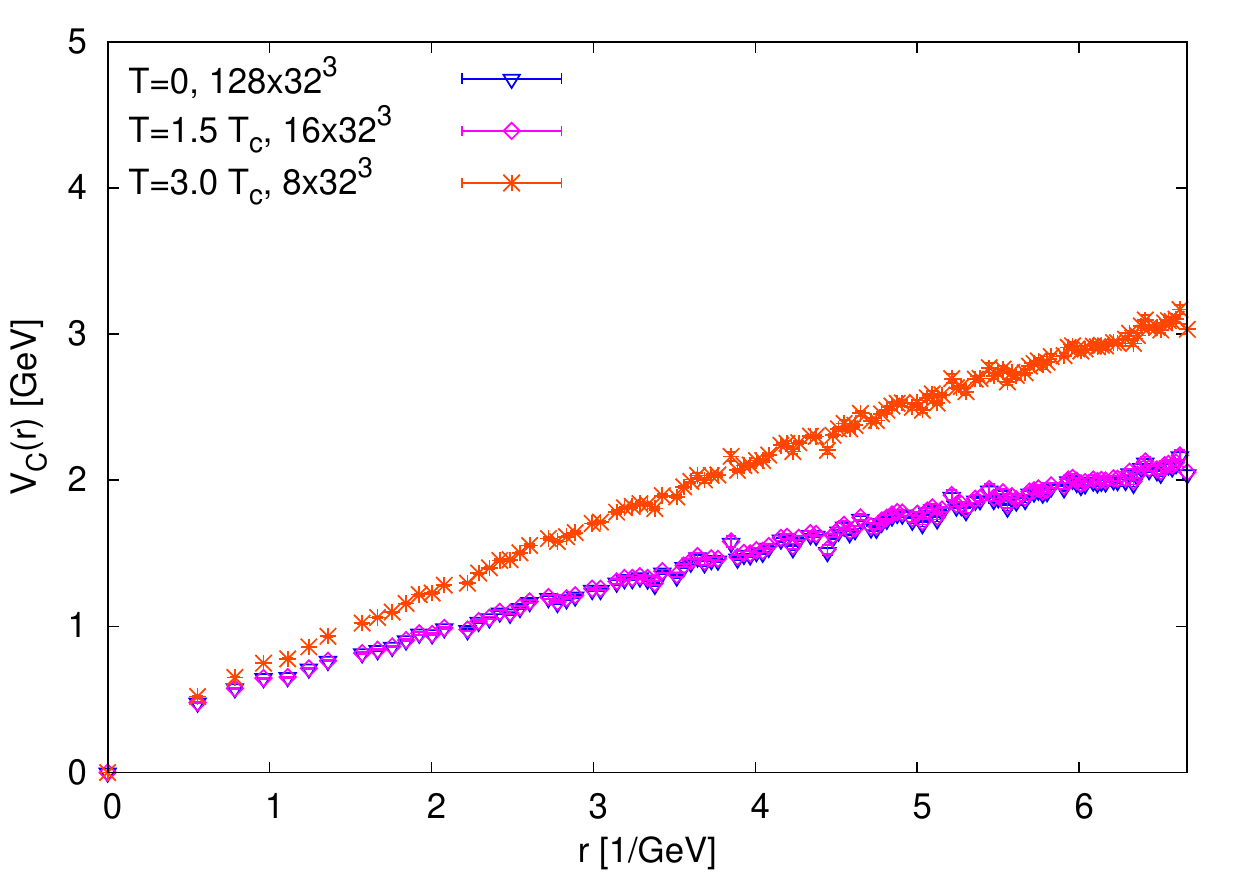}
	\caption{Coulomb potential from Eq.~(\ref{eq:correlator}) 
	for various temperatures at $\beta = 2.49$ 
        and anisotropy $\xi=4$. The slope of the curve at large distances
        is obviously constant at least up to $1.5 \, T_c$ and increases at $3 \, T_c$.}
	\label{fig3a}
\end{figure*}

As discussed in the introduction, a coherent picture of the Gribov-Zwanziger
confinement mechanism emerges from lattice Coulomb gauge investigations at 
$T=0$. As $T$ is increased, however, propagators do not seem to show a 
significant sensitivity to the deconfinement phase transition, as can
be seen in  Figs.~\ref{fig1}, \ref{fig2}, 
\ref{fig3}, \ref{fig3a} for the gluon propagator, the ghost
form factor and the Coulomb potential, respectively. Any deviation from the 
$T=0$ case starts well above $T =1.5\, T_c$. Also, the non-trivial infrared 
behaviour seems to be at first \emph{enhanced}, rather than decaying towards 
the perturbative
expectation.\footnote{The only exception might perhaps be he ghost form factor 
in Fig.~\ref{fig2}, whose IR exponent seems to decrease for $T \gg T_c$; such 
an exponent is however quite difficult to determine reliably, see
Ref.~\cite{Burgio:2012bk} for a critical discussion.}
In particular, the Coulomb string tension extracted from the 
Coulomb potential in Figs.~\ref{fig3}, \ref{fig3a} persists above the 
deconfinement phase transition, remaining constant up to $1.5 \, T_c$ and
increasing above it; see the figure captions for further details. 

These unexpected results were, in fact, the initial motivation for the present 
work. The fundamental puzzle is how it comes about that finite temperature 
correlators on the lattice decouple from the behaviour around $T_c$ 
expected from continuum investigations \cite{Heffner:2012sx}, 
while agreeing so well at $T=0$? Our working 
hypothesis is that the \emph{spatial} rather than the \emph{temporal} 
string tension underlies the finite temperature lattice Coulomb gauge 
dynamics. Indeed, the spatial string tension 
is known to persist and even rise above $T_c$, causing the strong correlations
expected in the quark-gluon plasma. We have therefore decided to investigate
this matter in more detail by going back to $T=0$ and controlling the string 
tension via the removal of (all or only spatial) center vortices in MC 
configurations.

\subsection{Vortex removal vs. Coulomb gauge}
\label{foo}

\subsubsection{String tension}
\label{secb1}

In a first step, we calculated the temporal and spatial string tensions 
through Creutz ratios \cite{Creutz:1980wj}, 
defined at distance $R$ as in Ref.~\cite{GonzalezArroyo:2012fx}: 
\begin{equation}
\chi(T+0.5,R+0.5) = -\log \frac{W(T+1,R+1)W(T,R)}{W(T+1,R)W(T,R+1)}\,.
\end{equation}
To reduce the statistical noise we used 5 steps of APE smearing 
\cite{Albanese:1987ds} with $\alpha = 0.5$ for all links, or only for 
the spatial links (if only spatial vortices were removed); such a procedure 
cannot of course be applied to the center projected links.

\begin{figure*}[htb]
	\center
	\includegraphics[width=0.99\columnwidth]{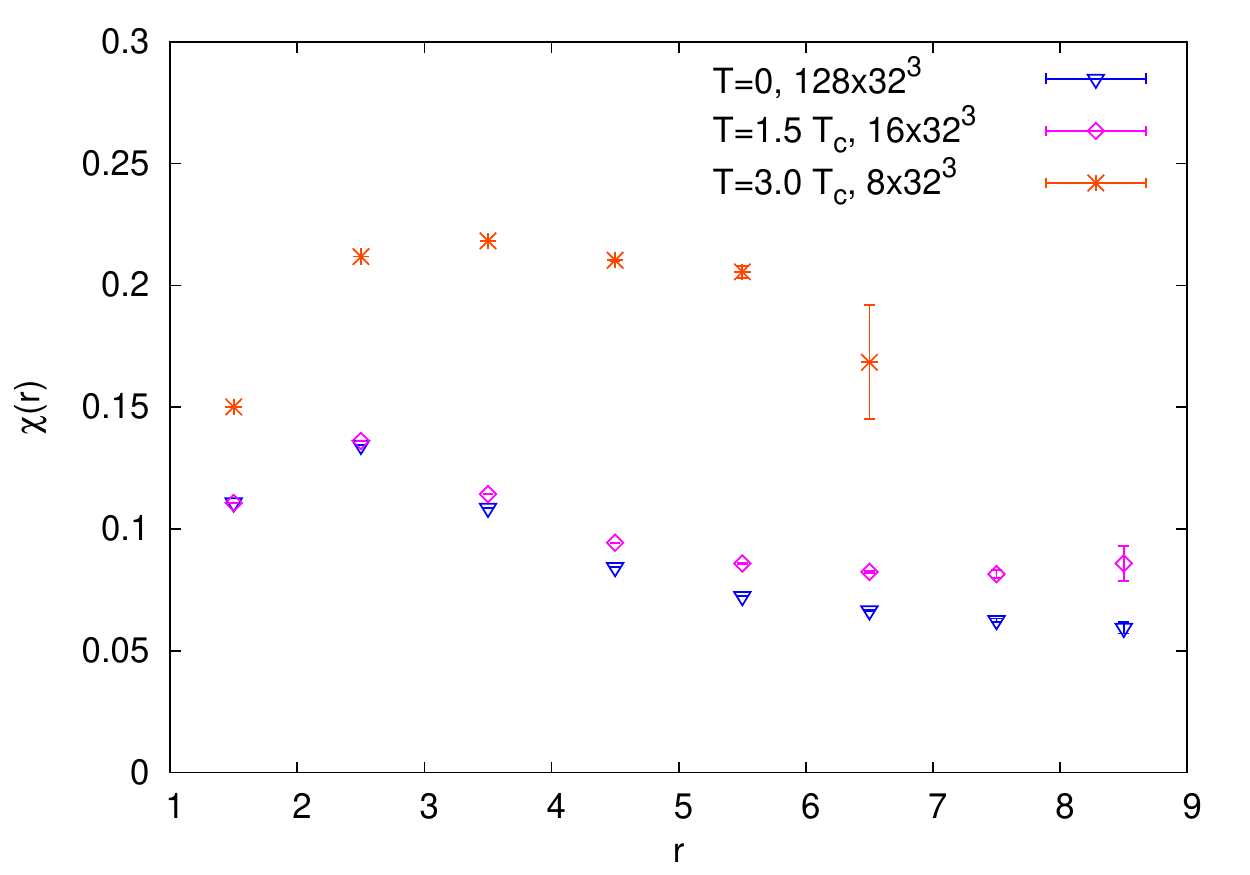}
	\caption{Creutz ratios for the spatial string tension
                 from the same configurations as in Figs.~\ref{fig3}, 
                 \ref{fig3a}; the dependence with the temperature
                  is the same. Data at $3 \, T_c$ 
                  for $r>7$ have been omitted, since the signal to noise 
                  ratio was very poor.}
	 \label{fig3b}
\end{figure*}

We first calculated the spatial string tension at zero and
finite temperatures to confirm that its dependence from the
temperature mimics the one we found for the Coulomb potential in 
Figs.~\ref{fig3}, \ref{fig3a}. In Fig.~\ref{fig3b} we show the spatial string
tension for exactly the same configurations from which 
Figs.~\ref{fig3}, \ref{fig3a} were calculated. The signal to noise ratio tends
to get worse as $T$ and $r$ rise; the data for $T=3 \, T_c$ above $r=7$
have been omitted, since their error-bars are orders of magnitude higher than
the reported data. Still, it is obvious from the plot that, within errors, the 
spatial string tensions hardly changes up to $1.5 \, T_c$, while it is
consistently higher at $3 \, T_c$.
\begin{figure*}[htb]
	\center
	\includegraphics[width=0.99\columnwidth]{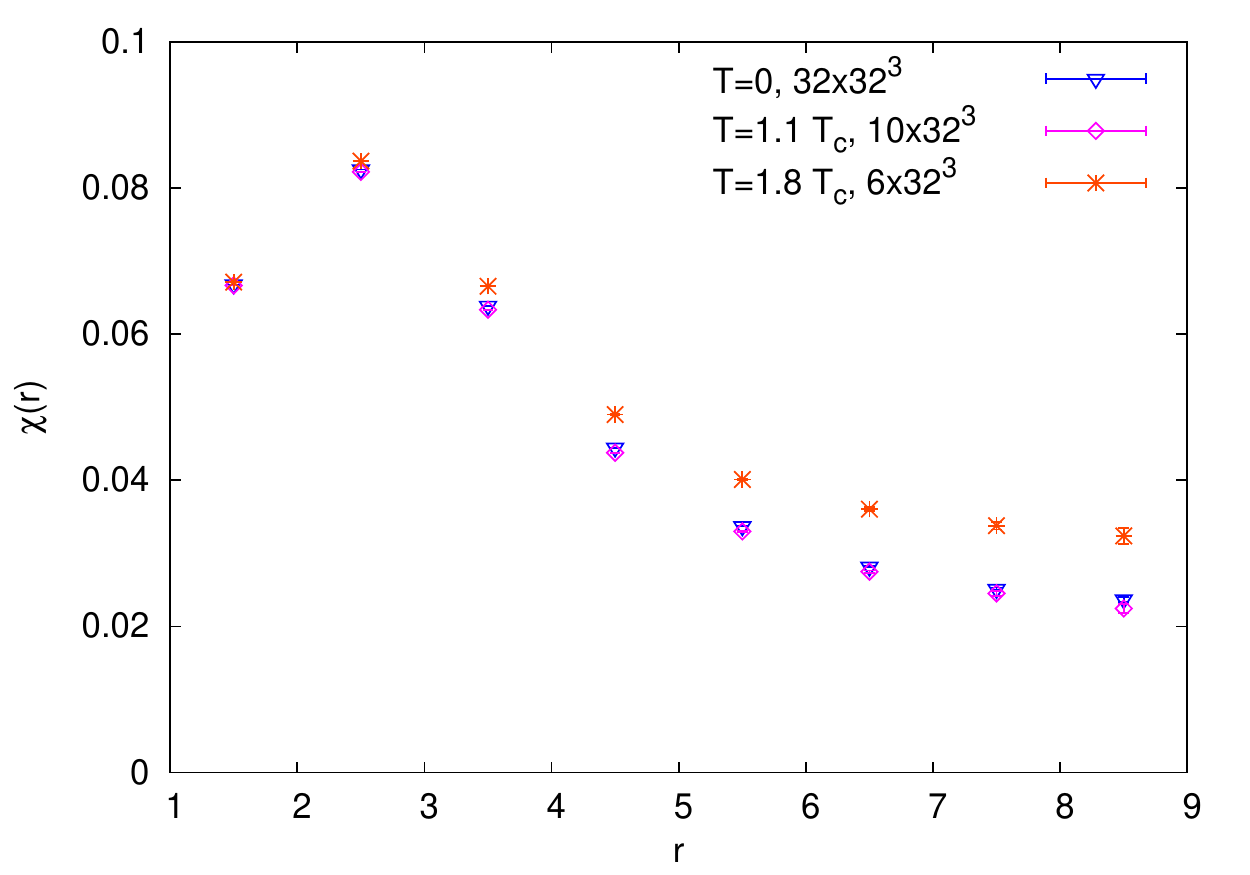}
	\caption{Spatial string tension at $\beta=2.6$ for
                 isotropic lattices up to a maximum temperature $1.8 \, T_c$.}
	 \label{fig3c}
\end{figure*}
As a cross check, we have calculated the spatial string tension for
a value of $\beta$ and at temperatures for which the signal to noise ratio
is known to be good; results are shown in Fig.~\ref{fig3c}. The $T=0$ and the 
$T= 1.1\,T_c$ data are indistinguishable within error-bars; on the other hand,
the $1.8\, T_c$ data extrapolate to a higher spatial string tension, consistent
with our expectations.

The above results show that we ought to be on a good path in assuming
that the Coulomb and the spatial string tensions are correlated, since both
behave semi-quantitatively in the same way as the temperature is increased.
We still however need to check the the spatial string tension is the
``cause'' for the Coulomb string tension. To achieve this, we will check
what happens to the Coulomb string tension after removing degrees of freedom
which are known to be strongly correlated to the confinement properties
of the theory, namely center vortices.

\begin{figure*}[htb]
	\center
	\includegraphics[width=0.99\columnwidth]{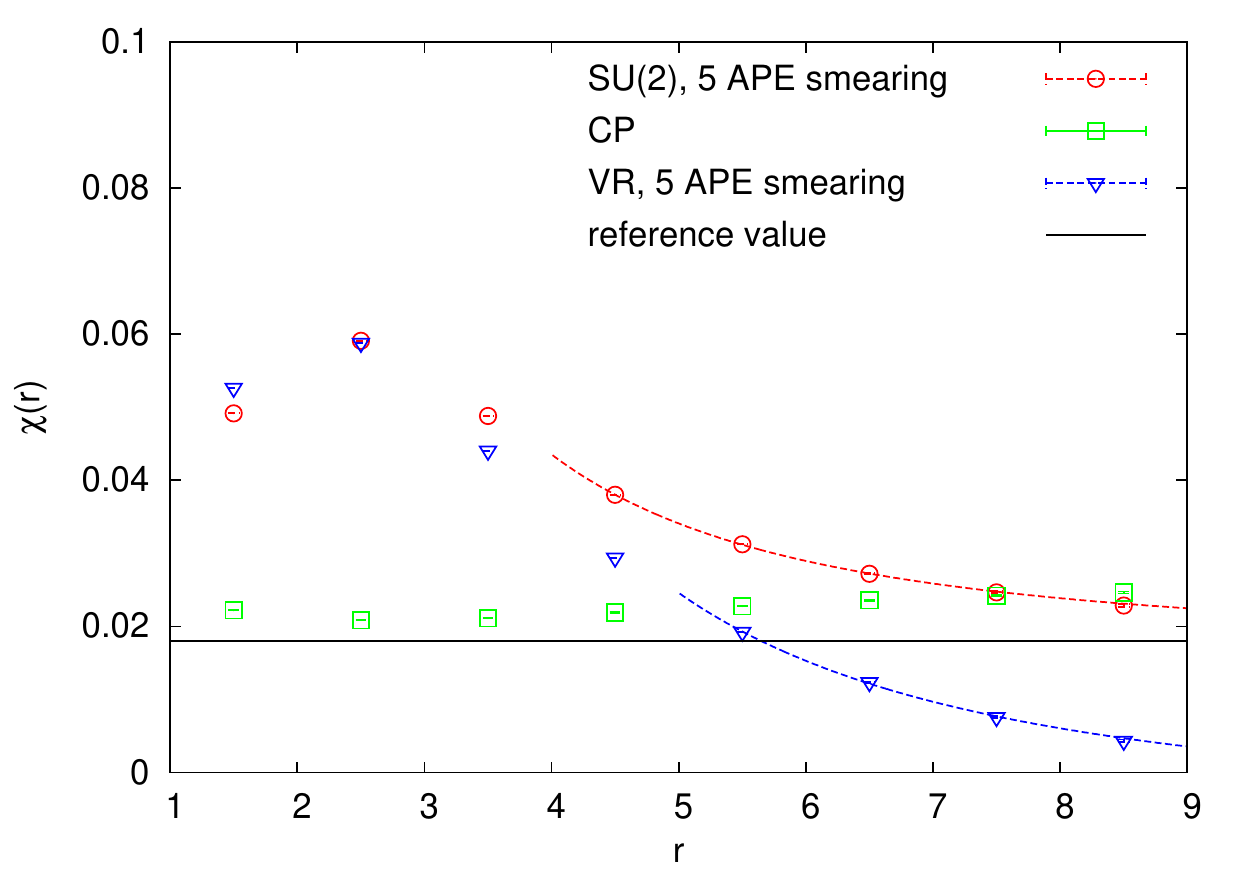}
	\caption{Creutz ratios for full center vortex projection (CP) and removal (VR)
	 at $\beta=2.6$. The reference value $\chi = 0.01915$ was taken from 
         Ref.~\cite{Bloch:2003sk}. Dotted lines are fits to the formula 
         $\sigma + \frac{2 k}{r^2}$, see 
         Ref.~\cite{GonzalezArroyo:2012fx} for further details.}
	 \label{fig4}
\end{figure*}

\begin{figure*}[htb]
	\center
	\includegraphics[width=0.99\columnwidth]{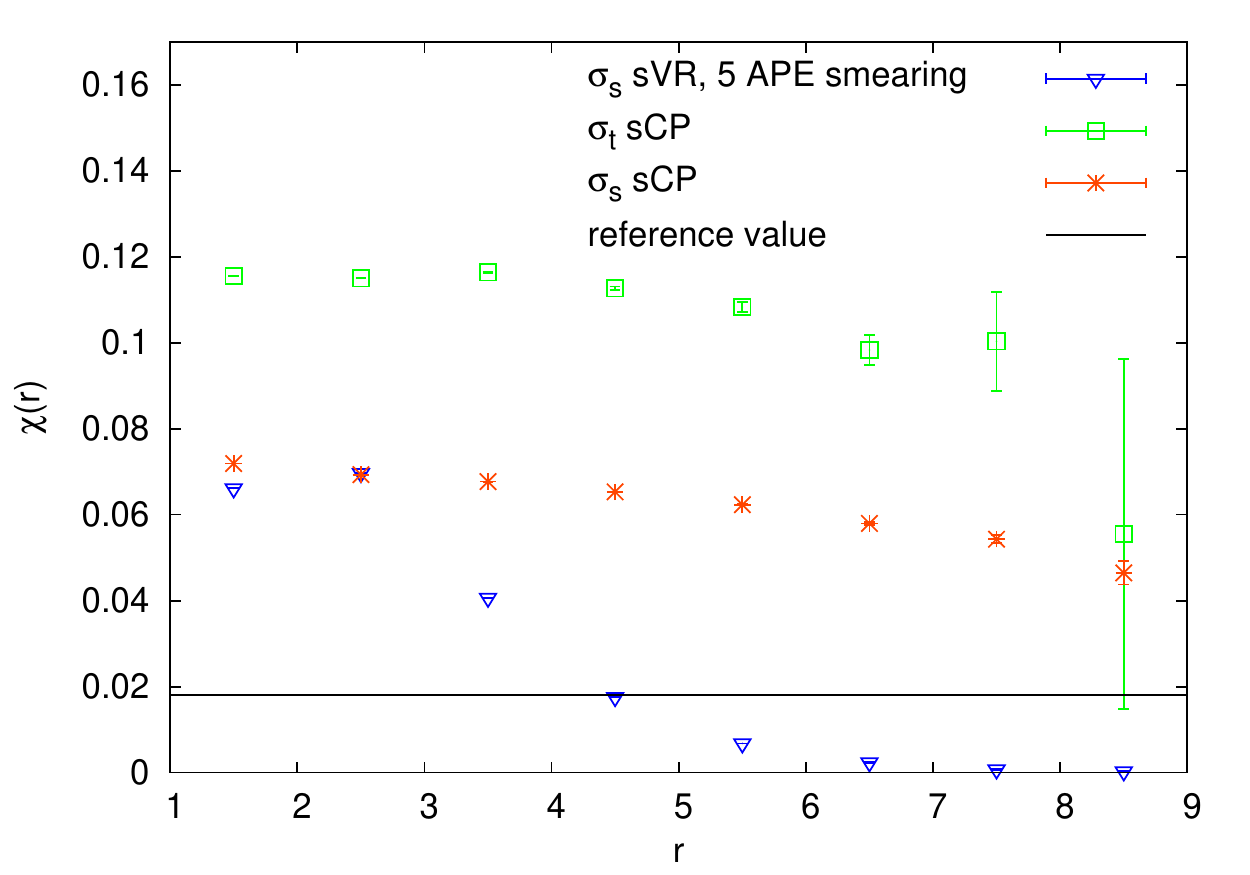}\\
	\caption{Creutz ratios for spatial center vortex projection and removal 
        at $\beta=2.6$, with the reference value taken from 
        Ref.~\cite{Bloch:2003sk}. {The symbols refer to the temporal 
        ($\sigma_t$) and spatial ($\sigma_s$) string tension in the spatially 
        center projected
        (sCP) and spatial vortex removed (sVR) ensemble, respectively.}
        }
	\label{fig5}
\end{figure*}

\begin{figure*}[htb]
	\center
	\includegraphics[width=0.99\columnwidth]{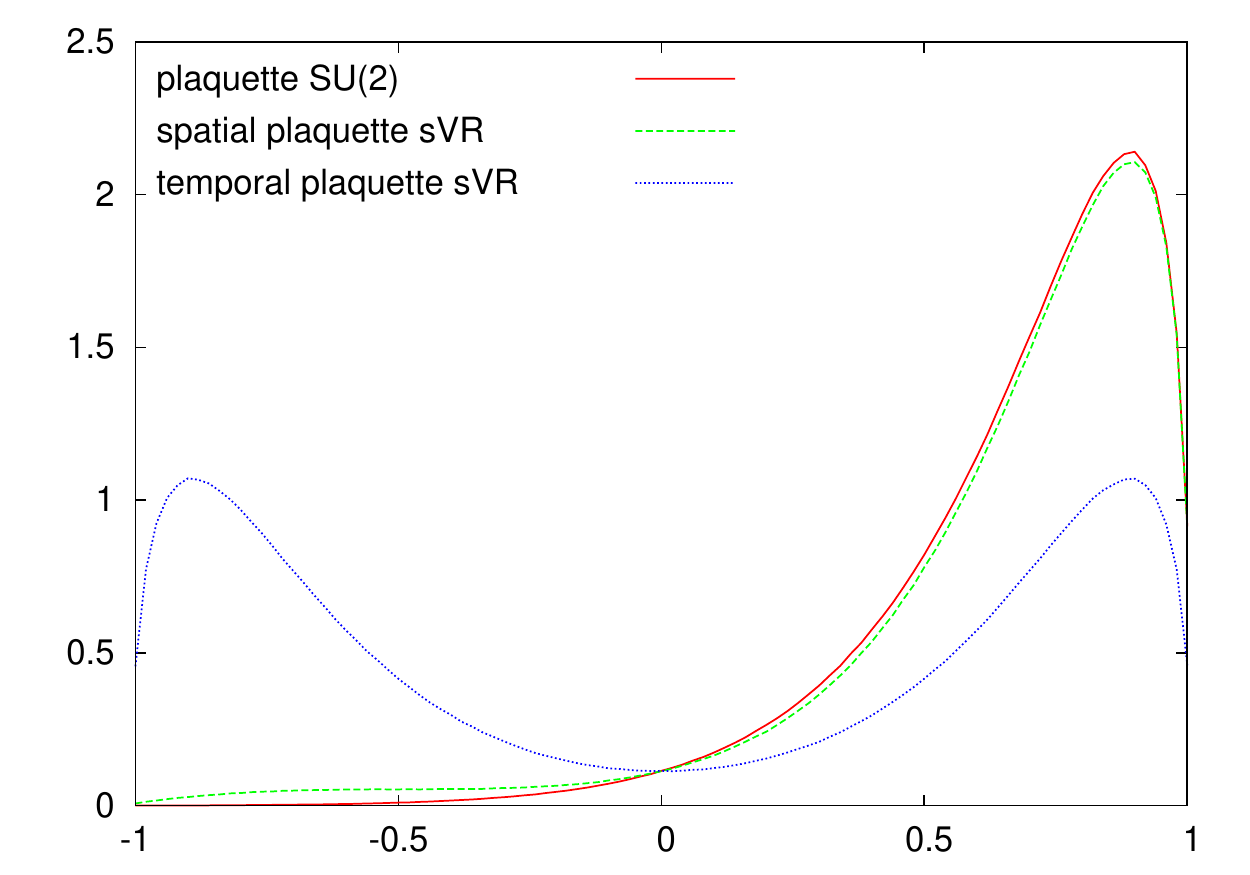}
	\caption{Histogram of plaquettes before and after spatial vortex 
        removal.}
	\label{fig6}
\end{figure*}

As a first check we verified that, as expected from the 
literature, the Wilson string tension drops to zero after full 
vortex removal {(VR)} 
and, conversely, keeps its $SU(2)$ value after full center projection {(CP)}
(see Fig.~\ref{fig4}). 
Next, we repeat this procedure (vortex removal and center projection) 
in the \emph{spatial} directions only. For the resulting configurations, 
it is necessary to distinguish between the temporal $\chi(T,R)$ and the spatial 
Creutz ratios $\chi(R_1,R_2)$. As expected, the spatial string tension 
$\sigma_s$ drops to zero after removing all spatial vortices, 
cf.~Fig.~\ref{fig5}. 
On the other hand, a direct 
measurement of the temporal string tension turns out to be impossible: as 
illustrated in the histogram in Fig.~\ref{fig6}, all 
space-time Wilson loops receive \emph{random} sign flips through spatial 
vortex removal; the signal to noise ratio becomes hopeless. However,
the \emph{temporal} string tension $\sigma_t$ measured from 
Polyakov loop correlators \emph{cannot} change under spatial center projection 
{(sCP)} or vortex removal {(sVR)}, since both procedures do not affect the 
temporal links from which the Polyakov lines are built. We can therefore
safely conclude that all spatial vortex removed configurations 
are still confining, exhibiting the exact same value for $\sigma_t$ as the 
original, gauge-unfixed ones.

The spatial projection, with or without vortex removal, can further 
introduce gauge noise in the temporal links if followed by a Coulomb gauge 
fixing. This makes a direct measurement 
of the temporal string tension through Creutz-ratios challenging, as can be 
seen from the large error bars arising at large distances $r$ 
in Fig.~\ref{fig5}. 
As can also be seen from this figure, both string tensions still exceed the 
asymptotic $SU(2)$ reference string tension at distances as large as 
$r \sim 9$, where they either have not yet reached a plateau (spatial) or are 
disappearing in statistical noise (temporal). We do not have a clear 
explanation for this slow convergence.

\subsubsection{Ghost form factor}
\begin{figure*}[htb]
	\center
	\includegraphics[width=0.99\columnwidth]{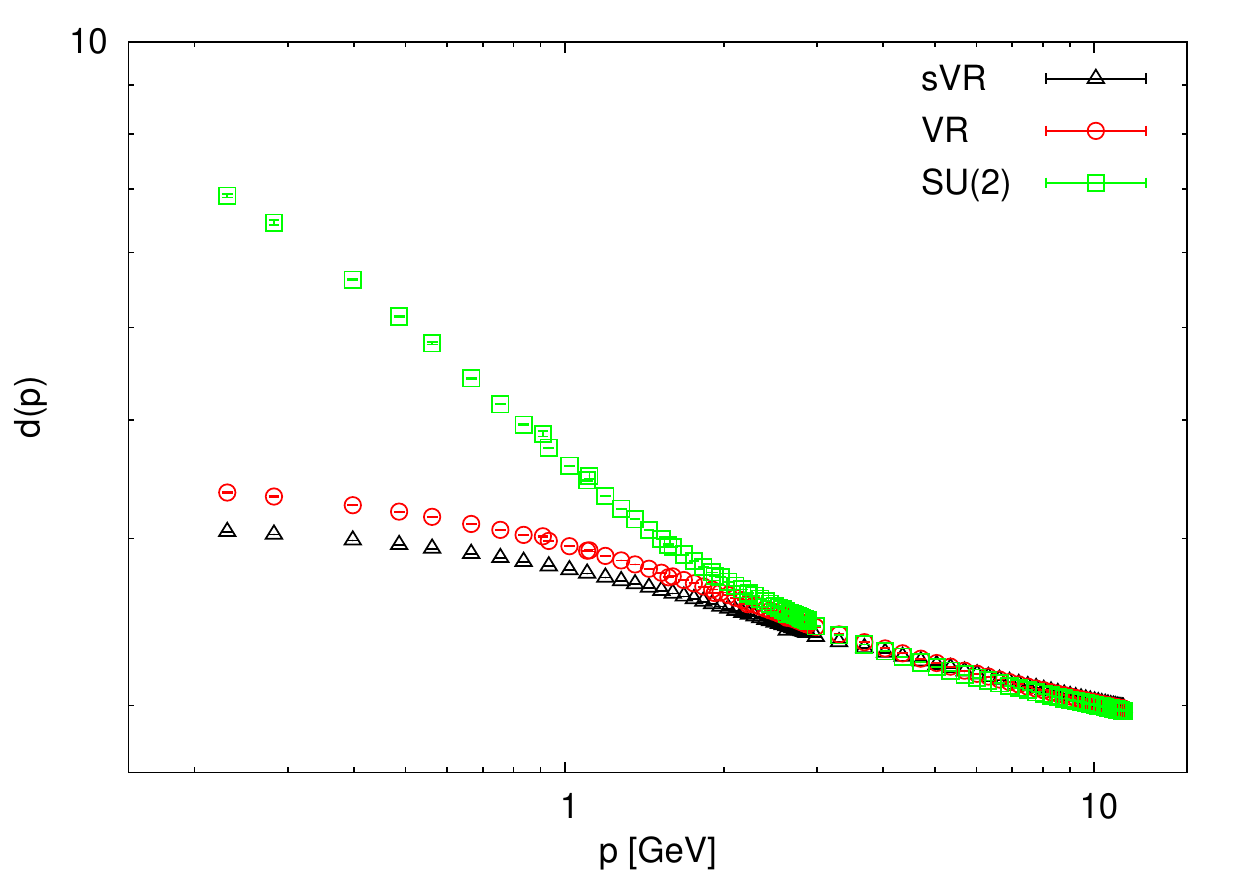}
	\caption{{Renormalized} ghost form factor before and after 
        vortex removal at $\beta=2.15$ and $\beta=2.60$. {The data was 
        multiplicatively renormalized at a reference scale of 
        $\mu = 3 \mathrm{GeV}$ whence the curves for the 
	two couplings $\beta$ fall on top of each other.}}
	\label{fig7}
\end{figure*}

\begin{figure*}[htb]
	\center
	\includegraphics[width=0.99\columnwidth]{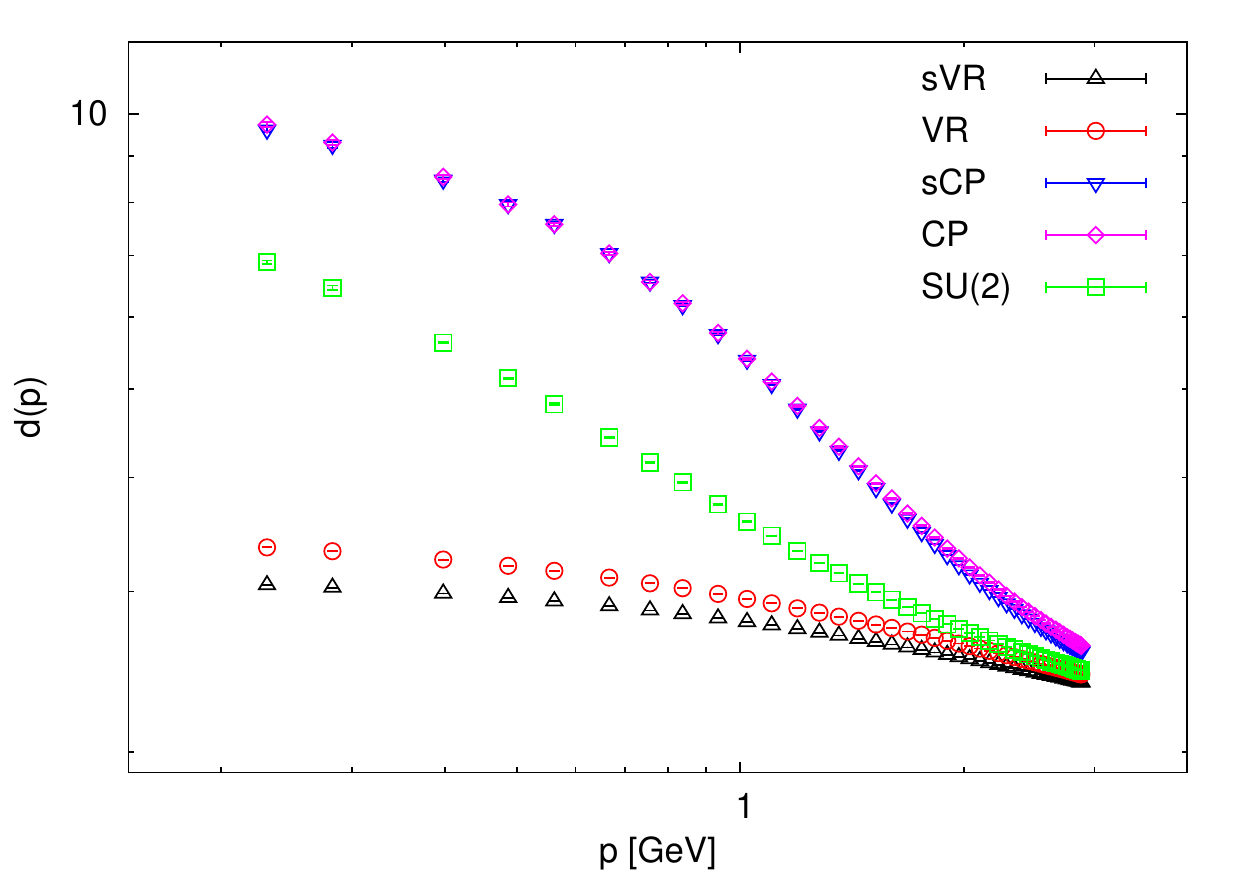}
	\caption{Ghost form factor after center projection (and restriction to 
        the non-zero subspace) at $\beta=2.15$.}
	\label{fig8}
\end{figure*}

From the results above, it is obvious that the MC configurations after spatial 
vortex removal still exhibit temporal confinement but no spatial confinement.
It is interesting to see how the Coulomb gauge correlators react to this change 
of physics in the underlying ensemble. As shown in Fig.~\ref{fig7},
the ghost form factor is no longer compatible with a power law in the deep 
infrared, both after full and spatial vortex removal. As for the center 
projected configurations, a naive computation of ghost propagator is 
ill-defined 
because the Faddeev-Popov (FP) operator acquires additional zero modes from 
the center vortices which sit directly on the Gribov 
horizon.\footnote{The $N_c^2-1$ 
\emph{constant} zero modes are easy to take care of by restricting the 
calculation to momenta $p \neq 0$.} It is, however, possible to invert the 
FP operator in the 
subspace orthogonal to the kernel, which thus gives the subleading
contributions to $d(\vec{p})$. The result is shown in 
Fig.~\ref{fig8}, 
where we observe an enhancement in the mid-momentum regime and a 
suppression in the deep infra-red as compared to the unprojected Coulomb gauge. 
From these investigations, it is clear that the infra-red enhancement of the 
original Coulomb gauge form factor is in fact tied to the spatial string 
tension, 
as elimination of the latter leads to to a dramatic suppression of the former 
in the IR.

\subsubsection{Coulomb potential}
\begin{figure*}[htb]
	\center
	\includegraphics[width=0.99\columnwidth]{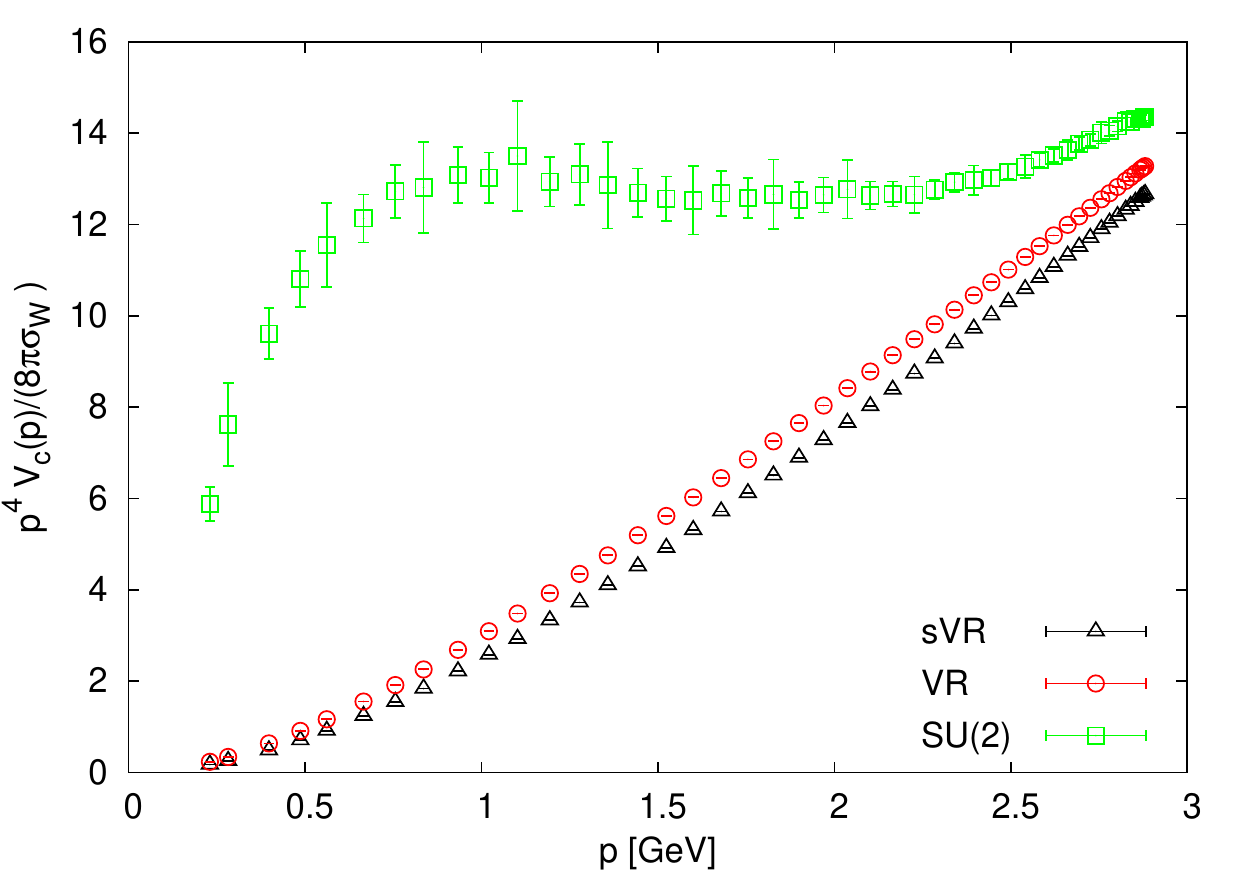}
	\caption{{Standard} Coulomb potential Eq.~(\ref{eq:coulombpotential}) 
	 {in momentum space}, normalized such that the 
	intercept at $p \to 0$ yields $8 \pi \sigma_C$.
	{The relatively sharp drop in the deep infrared for the full $SU(2)$ 
	data made it necessary to take a fairly small coupling $\beta = 2.15$ 
	to (roughly) estimate the intercept.}}
	\label{fig9}
\end{figure*}

The extrapolation of the Coulomb string tension $\sigma_C$ from the potential 
Eq.~\eqref{eq:coulombpotential} is possible but suffers from 
large uncertainties for a variety of reasons. Estimates were given in
Refs.~\cite{Greensite:2003xf,Greensite:2004ke,Voigt:2008rr,Nakagawa:2008zza,%
Nakagawa:2010eh,Burgio:2012bk}. 
We follow the convention in the literature and plot in 
Fig.~\ref{fig9} the ratio $p^4 V_C(p) / (8 \pi \sigma_W)$, 
{which yields $\sigma_C / \sigma_W$ in the limit $p \to 0$, 
cf.~the remark after Eq.~(\ref{eq:coulombpotential}).}
As can be clearly seen from the plot, the Coulomb string tension $\sigma_C$ 
disappears after both full and spatial vortex removal. Since the latter case 
still contains the full temporal string tension, as discussed in 
Sec.~\ref{secb1},
it is clear that the 
definition of the Coulomb string tension through 
Eq.~\eqref{eq:coulombpotential} must be directly related to the spatial string 
tension. 

\begin{figure*}[htb]
	\center
	\includegraphics[width=0.99\columnwidth]{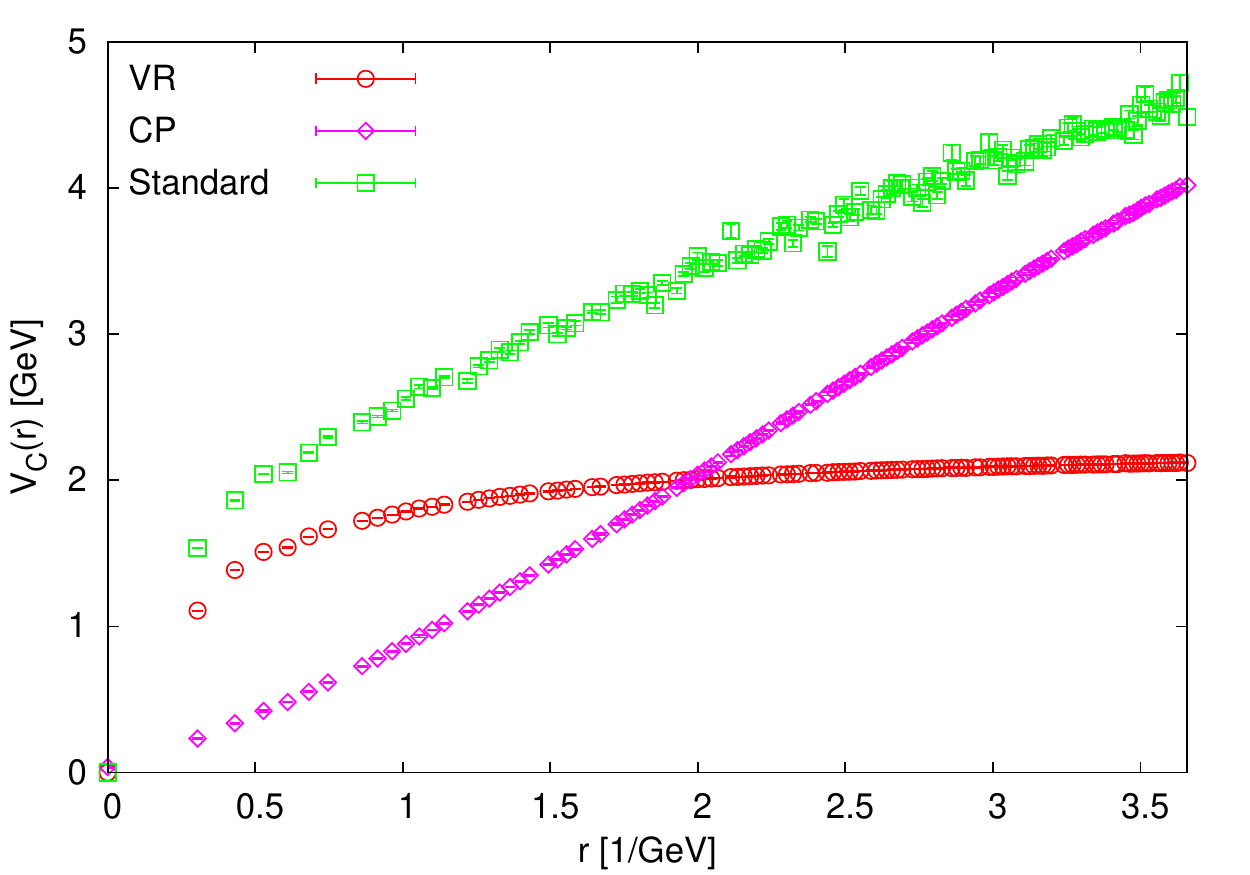}
	\caption{Coulomb potential from Eq.~\eqref{eq:correlator}
	{in position space. Linear behaviour sets in at moderate distances $r$, 
	so that a fairly large coupling $\beta = 2.60$ could be afforded to 
        minimize discretization effects. See the discussion at the end of 
        section \ref{foo} for further details.}}
	\label{fig10}
\end{figure*}

It is interesting now to consider Eq.~\eqref{eq:correlator}, our 
alternative 
definition of the Coulomb potential. From Ref.~\cite{Greensite:2003xf} this 
is known to allow for a better extrapolation of the Coulomb string tension 
while still vanishing after full vortex removal, cf. Fig.~\ref{fig10}, where
it has been plotted together with its full center-projected and
full vortex removed counterpart.
\begin{figure*}[htb]
	\center
	\includegraphics[width=0.99\columnwidth]{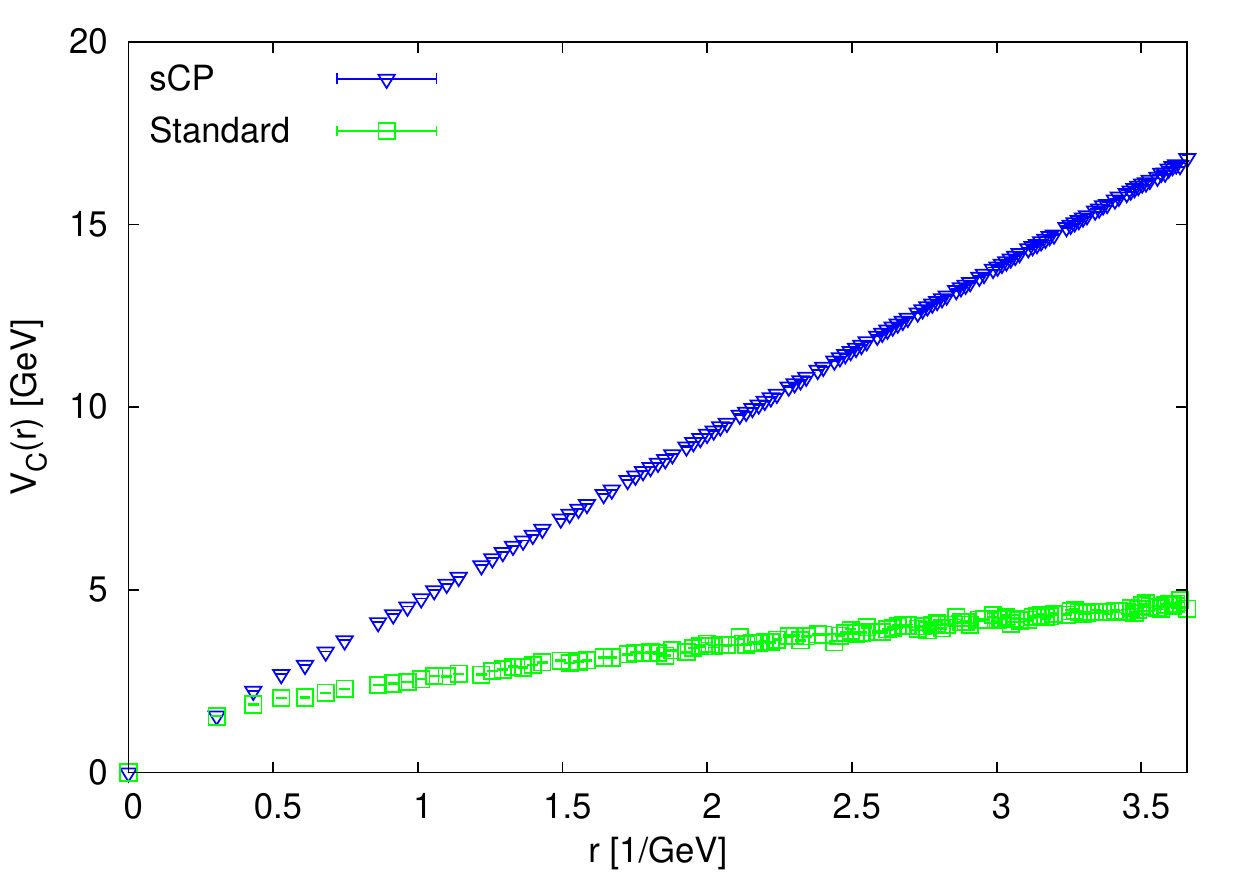}
	\caption{Coulomb potential from Eq.~\eqref{eq:correlator}
	{in position space.} Both the spatial center projected {\it and}
        the spatial vortex removed correlators are compatible with a
        linear rising potential, although the former overestimates the
        string tension by a large factor.}
	\label{fig11}
\end{figure*}
On the other hand, the spatial vortex removed correlators must remain 
unchanged since 
Eq.~\eqref{eq:correlator} employs the temporal links $U_0(\vec{x})$ only. Remarkably, 
its spatial center projected counterpart still rises linearly,
as it can be seen in Fig.~\ref{fig11}. 
It is thus clear that the definition given in Eq.~\eqref{eq:correlator} 
of the Coulomb potential is 
indeed sensitive to both the temporal and the spatial string tension, 
as it still raises linearly both after spatial center projection and spatial
vortex removal. This is not in contrast with 
the result found in Sec.~\ref{sec:fT}: above the phase transition
the temporal string tension vanishes, thus leaving only the spatial
string tension to affect the $U_0(\vec{x})$ correlators at high temperatures.
Therefore, when either the temporal or the spatial string tension
disappears, as above the deconfinement
transiton (temporal), see Fig.~\ref{fig3a},
or after spatial vortex removal (spatial), see Fig.~\ref{fig11}, 
Eq.~\eqref{eq:correlator} will still give a linear rising potential. 
Only when both are eliminated via full vortex removal, see Fig.~\ref{fig10}
the $U_0(\vec{x})$ correlator becomes asimptotically flat.

Eq.~\eqref{eq:correlator} is therefore some kind of hybrid 
definition, sensitive to the confining string tension but, because of its 
``distance'' to the Hamiltonian formulation due to being defined on a single 
time slice, still not sufficient to have overlap with all the excited states. 
The very large value it assumes after 
spatial center gauge fixing and projection (see Fig.~\ref{fig11}) is likely 
a related phenomenon, since a mixing of degrees of freedom obviously occurs.
Modifications of Eq.~\eqref{eq:correlator} might offer better results, 
since correlators of longer open Polyakov lines could turn out to be closer 
to the finite temperature dynamics in Coulomb gauge. However, as the length of 
the line increases, the relationship with the original Coulomb potential 
becomes obfuscated. Thus, a static Coulomb gauge observable that can detect the 
deconfinement phase transition on the lattice remains somewhat elusive. 

\section{Conclusions}

In this paper we have investigated the relationship between spatial and
Coulomb string tension as measured through the standard lattice definition of 
Coulomb gauge correlators. Such an observables are made out of the space-like 
links at a fixed time slice and, as we have seen, can only be used for 
investigations at $T=0$. As the temperature increases, temporal and spatial 
string tension decouple and we find that the dynamics of static Coulomb gauge 
observables are clearly dominated by the latter and not the former. This 
explains why the Coulomb string tension from 
Eq.~(\ref{eq:coulombpotential}) persists above $T_c$, and the low-order lattice
Green's functions do not react to the loss of the temporal string tension at
and above $T_c$. 

The alternative 
definition of the Coulomb potential pioneered in Ref.~\cite{Marinari:1992kh},
on the other hand, turns out to detect \emph{both} the temporal 
and the spatial string tension, but
it still cannot fully resolve the deconfinement phase transition, as this would
probably require longer lines with temporal extensions comparable to the first 
excited states of the theory \cite{Greensite:2003xf}. Such an observables 
can, however, no longer be easily related to the static Coulomb potential.
More refined lattice observables are clearly necessary, and they may be 
tested with the methods layed out in this paper.

\begin{acknowledgments}
This work was partially supported by the Deutsche Forschungsgemeinschaft
under contract DFG-Re 856/9-1.
H.V. wishes to thank the Evangelisches Studienwerk Villigst for financial 
support.
 
\end{acknowledgments}

\bibliographystyle{apsrev4-1}
\bibliography{centerpaper}

\begin{thebibliography}{81}%
\makeatletter
\providecommand \@ifxundefined [1]{%
 \@ifx{#1\undefined}
}%
\providecommand \@ifnum [1]{%
 \ifnum #1\expandafter \@firstoftwo
 \else \expandafter \@secondoftwo
 \fi
}%
\providecommand \@ifx [1]{%
 \ifx #1\expandafter \@firstoftwo
 \else \expandafter \@secondoftwo
 \fi
}%
\providecommand \natexlab [1]{#1}%
\providecommand \enquote  [1]{``#1''}%
\providecommand \bibnamefont  [1]{#1}%
\providecommand \bibfnamefont [1]{#1}%
\providecommand \citenamefont [1]{#1}%
\providecommand \href@noop [0]{\@secondoftwo}%
\providecommand \href [0]{\begingroup \@sanitize@url \@href}%
\providecommand \@href[1]{\@@startlink{#1}\@@href}%
\providecommand \@@href[1]{\endgroup#1\@@endlink}%
\providecommand \@sanitize@url [0]{\catcode `\\12\catcode `\$12\catcode
  `\&12\catcode `\#12\catcode `\^12\catcode `\_12\catcode `\%12\relax}%
\providecommand \@@startlink[1]{}%
\providecommand \@@endlink[0]{}%
\providecommand \url  [0]{\begingroup\@sanitize@url \@url }%
\providecommand \@url [1]{\endgroup\@href {#1}{\urlprefix }}%
\providecommand \urlprefix  [0]{URL }%
\providecommand \Eprint [0]{\href }%
\providecommand \doibase [0]{http://dx.doi.org/}%
\providecommand \selectlanguage [0]{\@gobble}%
\providecommand \bibinfo  [0]{\@secondoftwo}%
\providecommand \bibfield  [0]{\@secondoftwo}%
\providecommand \translation [1]{[#1]}%
\providecommand \BibitemOpen [0]{}%
\providecommand \bibitemStop [0]{}%
\providecommand \bibitemNoStop [0]{.\EOS\space}%
\providecommand \EOS [0]{\spacefactor3000\relax}%
\providecommand \BibitemShut  [1]{\csname bibitem#1\endcsname}%
\let\auto@bib@innerbib\@empty
\bibitem [{\citenamefont {Zwanziger}(2003)}]{Zwanziger:2002sh}%
  \BibitemOpen
  \bibfield  {author} {\bibinfo {author} {\bibfnamefont {D.}~\bibnamefont
  {Zwanziger}},\ }\href {\doibase 10.1103/PhysRevLett.90.102001} {\bibfield
  {journal} {\bibinfo  {journal} {Phys.Rev.Lett.}\ }\textbf {\bibinfo {volume}
  {90}},\ \bibinfo {pages} {102001} (\bibinfo {year} {2003})},\ \Eprint
  {http://arxiv.org/abs/hep-lat/0209105} {arXiv:hep-lat/0209105 [hep-lat]}
  \BibitemShut {NoStop}%
\bibitem [{\citenamefont {Jackiw}\ \emph {et~al.}(1978)\citenamefont {Jackiw},
  \citenamefont {Muzinich},\ and\ \citenamefont {Rebbi}}]{Jackiw:1977ng}%
  \BibitemOpen
  \bibfield  {author} {\bibinfo {author} {\bibfnamefont {R.}~\bibnamefont
  {Jackiw}}, \bibinfo {author} {\bibfnamefont {I.}~\bibnamefont {Muzinich}}, \
  and\ \bibinfo {author} {\bibfnamefont {C.}~\bibnamefont {Rebbi}},\ }\href
  {\doibase 10.1103/PhysRevD.17.1576} {\bibfield  {journal} {\bibinfo
  {journal} {Phys. Rev.}\ }\textbf {\bibinfo {volume} {D17}},\ \bibinfo {pages}
  {1576} (\bibinfo {year} {1978})}\BibitemShut {NoStop}%
\bibitem [{\citenamefont {Schutte}(1985)}]{Schutte:1985sd}%
  \BibitemOpen
  \bibfield  {author} {\bibinfo {author} {\bibfnamefont {D.}~\bibnamefont
  {Schutte}},\ }\href {\doibase 10.1103/PhysRevD.31.810} {\bibfield  {journal}
  {\bibinfo  {journal} {Phys.Rev.}\ }\textbf {\bibinfo {volume} {D31}},\
  \bibinfo {pages} {810} (\bibinfo {year} {1985})}\BibitemShut {NoStop}%
\bibitem [{\citenamefont {Szczepaniak}\ and\ \citenamefont
  {Swanson}(2002)}]{Szczepaniak:2001rg}%
  \BibitemOpen
  \bibfield  {author} {\bibinfo {author} {\bibfnamefont {A.~P.}\ \bibnamefont
  {Szczepaniak}}\ and\ \bibinfo {author} {\bibfnamefont {E.~S.}\ \bibnamefont
  {Swanson}},\ }\href {\doibase 10.1103/PhysRevD.65.025012} {\bibfield
  {journal} {\bibinfo  {journal} {Phys. Rev.}\ }\textbf {\bibinfo {volume}
  {D65}},\ \bibinfo {pages} {025012} (\bibinfo {year} {2002})},\ \Eprint
  {http://arxiv.org/abs/hep-ph/0107078} {arXiv:hep-ph/0107078} \BibitemShut
  {NoStop}%
\bibitem [{\citenamefont {Feuchter}\ and\ \citenamefont
  {Reinhardt}(2004)}]{Feuchter:2004mk}%
  \BibitemOpen
  \bibfield  {author} {\bibinfo {author} {\bibfnamefont {C.}~\bibnamefont
  {Feuchter}}\ and\ \bibinfo {author} {\bibfnamefont {H.}~\bibnamefont
  {Reinhardt}},\ }\href {\doibase 10.1103/PhysRevD.70.105021} {\bibfield
  {journal} {\bibinfo  {journal} {Phys. Rev.}\ }\textbf {\bibinfo {volume}
  {D70}},\ \bibinfo {pages} {105021} (\bibinfo {year} {2004})},\ \Eprint
  {http://arxiv.org/abs/hep-th/0408236} {arXiv:hep-th/0408236} \BibitemShut
  {NoStop}%
\bibitem [{\citenamefont {Schleifenbaum}\ \emph {et~al.}(2006)\citenamefont
  {Schleifenbaum}, \citenamefont {Leder},\ and\ \citenamefont
  {Reinhardt}}]{Schleifenbaum:2006bq}%
  \BibitemOpen
  \bibfield  {author} {\bibinfo {author} {\bibfnamefont {W.}~\bibnamefont
  {Schleifenbaum}}, \bibinfo {author} {\bibfnamefont {M.}~\bibnamefont
  {Leder}}, \ and\ \bibinfo {author} {\bibfnamefont {H.}~\bibnamefont
  {Reinhardt}},\ }\href {\doibase 10.1103/PhysRevD.73.125019} {\bibfield
  {journal} {\bibinfo  {journal} {Phys. Rev.}\ }\textbf {\bibinfo {volume}
  {D73}},\ \bibinfo {pages} {125019} (\bibinfo {year} {2006})},\ \Eprint
  {http://arxiv.org/abs/hep-th/0605115} {arXiv:hep-th/0605115} \BibitemShut
  {NoStop}%
\bibitem [{\citenamefont {Epple}\ \emph {et~al.}(2007)\citenamefont {Epple},
  \citenamefont {Reinhardt},\ and\ \citenamefont
  {Schleifenbaum}}]{Epple:2006hv}%
  \BibitemOpen
  \bibfield  {author} {\bibinfo {author} {\bibfnamefont {D.}~\bibnamefont
  {Epple}}, \bibinfo {author} {\bibfnamefont {H.}~\bibnamefont {Reinhardt}}, \
  and\ \bibinfo {author} {\bibfnamefont {W.}~\bibnamefont {Schleifenbaum}},\
  }\href {\doibase 10.1103/PhysRevD.75.045011} {\bibfield  {journal} {\bibinfo
  {journal} {Phys. Rev.}\ }\textbf {\bibinfo {volume} {D75}},\ \bibinfo {pages}
  {045011} (\bibinfo {year} {2007})},\ \Eprint
  {http://arxiv.org/abs/hep-th/0612241} {arXiv:hep-th/0612241} \BibitemShut
  {NoStop}%
\bibitem [{\citenamefont {Epple}\ \emph {et~al.}(2008)\citenamefont {Epple},
  \citenamefont {Reinhardt}, \citenamefont {Schleifenbaum},\ and\ \citenamefont
  {Szczepaniak}}]{Epple:2007ut}%
  \BibitemOpen
  \bibfield  {author} {\bibinfo {author} {\bibfnamefont {D.}~\bibnamefont
  {Epple}}, \bibinfo {author} {\bibfnamefont {H.}~\bibnamefont {Reinhardt}},
  \bibinfo {author} {\bibfnamefont {W.}~\bibnamefont {Schleifenbaum}}, \ and\
  \bibinfo {author} {\bibfnamefont {A.~P.}\ \bibnamefont {Szczepaniak}},\
  }\href {\doibase 10.1103/PhysRevD.77.085007} {\bibfield  {journal} {\bibinfo
  {journal} {Phys. Rev.}\ }\textbf {\bibinfo {volume} {D77}},\ \bibinfo {pages}
  {085007} (\bibinfo {year} {2008})},\ \Eprint {http://arxiv.org/abs/0712.3694}
  {arXiv:0712.3694 [hep-th]} \BibitemShut {NoStop}%
\bibitem [{\citenamefont {Feuchter}\ and\ \citenamefont
  {Reinhardt}(2008)}]{Feuchter:2007mq}%
  \BibitemOpen
  \bibfield  {author} {\bibinfo {author} {\bibfnamefont {C.}~\bibnamefont
  {Feuchter}}\ and\ \bibinfo {author} {\bibfnamefont {H.}~\bibnamefont
  {Reinhardt}},\ }\href {\doibase 10.1103/PhysRevD.77.085023} {\bibfield
  {journal} {\bibinfo  {journal} {Phys. Rev.}\ }\textbf {\bibinfo {volume}
  {D77}},\ \bibinfo {pages} {085023} (\bibinfo {year} {2008})},\ \Eprint
  {http://arxiv.org/abs/0711.2452} {arXiv:0711.2452 [hep-th]} \BibitemShut
  {NoStop}%
\bibitem [{\citenamefont {Reinhardt}\ and\ \citenamefont
  {Schleifenbaum}(2009)}]{Reinhardt:2008ij}%
  \BibitemOpen
  \bibfield  {author} {\bibinfo {author} {\bibfnamefont {H.}~\bibnamefont
  {Reinhardt}}\ and\ \bibinfo {author} {\bibfnamefont {W.}~\bibnamefont
  {Schleifenbaum}},\ }\href {\doibase 10.1016/j.aop.2008.09.005} {\bibfield
  {journal} {\bibinfo  {journal} {Annals Phys.}\ }\textbf {\bibinfo {volume}
  {324}},\ \bibinfo {pages} {735} (\bibinfo {year} {2009})},\ \Eprint
  {http://arxiv.org/abs/0809.1764} {arXiv:0809.1764 [hep-th]} \BibitemShut
  {NoStop}%
\bibitem [{\citenamefont {Reinhardt}(2008)}]{Reinhardt:2008ek}%
  \BibitemOpen
  \bibfield  {author} {\bibinfo {author} {\bibfnamefont {H.}~\bibnamefont
  {Reinhardt}},\ }\href {\doibase 10.1103/PhysRevLett.101.061602} {\bibfield
  {journal} {\bibinfo  {journal} {Phys. Rev. Lett.}\ }\textbf {\bibinfo
  {volume} {101}},\ \bibinfo {pages} {061602} (\bibinfo {year} {2008})},\
  \Eprint {http://arxiv.org/abs/0803.0504} {arXiv:0803.0504 [hep-th]}
  \BibitemShut {NoStop}%
\bibitem [{\citenamefont {Campagnari}\ and\ \citenamefont
  {Reinhardt}(2010)}]{Campagnari:2010wc}%
  \BibitemOpen
  \bibfield  {author} {\bibinfo {author} {\bibfnamefont {D.~R.}\ \bibnamefont
  {Campagnari}}\ and\ \bibinfo {author} {\bibfnamefont {H.}~\bibnamefont
  {Reinhardt}},\ }\href {\doibase 10.1103/PhysRevD.82.105021} {\bibfield
  {journal} {\bibinfo  {journal} {Phys.Rev.}\ }\textbf {\bibinfo {volume}
  {D82}},\ \bibinfo {pages} {105021} (\bibinfo {year} {2010})},\ \Eprint
  {http://arxiv.org/abs/1009.4599} {arXiv:1009.4599 [hep-th]} \BibitemShut
  {NoStop}%
\bibitem [{\citenamefont {Pak}\ and\ \citenamefont
  {Reinhardt}(2012)}]{Pak:2011wu}%
  \BibitemOpen
  \bibfield  {author} {\bibinfo {author} {\bibfnamefont {M.}~\bibnamefont
  {Pak}}\ and\ \bibinfo {author} {\bibfnamefont {H.}~\bibnamefont
  {Reinhardt}},\ }\href {\doibase 10.1016/j.physletb.2012.01.006} {\bibfield
  {journal} {\bibinfo  {journal} {Physics Letters B}\ }\textbf {\bibinfo
  {volume} {707}},\ \bibinfo {pages} {566 } (\bibinfo {year} {2012})},\ \Eprint
  {http://arxiv.org/abs/1107.5263} {arXiv:1107.5263 [hep-ph]} \BibitemShut
  {NoStop}%
\bibitem [{\citenamefont {Campagnari}\ and\ \citenamefont
  {Reinhardt}(2012)}]{Campagnari:2011bk}%
  \BibitemOpen
  \bibfield  {author} {\bibinfo {author} {\bibfnamefont {D.~R.}\ \bibnamefont
  {Campagnari}}\ and\ \bibinfo {author} {\bibfnamefont {H.}~\bibnamefont
  {Reinhardt}},\ }\href {\doibase 10.1016/j.physletb.2011.12.024} {\bibfield
  {journal} {\bibinfo  {journal} {Phys.Lett.}\ }\textbf {\bibinfo {volume}
  {B707}},\ \bibinfo {pages} {216} (\bibinfo {year} {2012})},\ \Eprint
  {http://arxiv.org/abs/1111.5476} {arXiv:1111.5476 [hep-th]} \BibitemShut
  {NoStop}%
\bibitem [{\citenamefont {Reinhardt}\ \emph {et~al.}(2011)\citenamefont
  {Reinhardt}, \citenamefont {Campagnari},\ and\ \citenamefont
  {Szczepaniak}}]{Reinhardt:2011hq}%
  \BibitemOpen
  \bibfield  {author} {\bibinfo {author} {\bibfnamefont {H.}~\bibnamefont
  {Reinhardt}}, \bibinfo {author} {\bibfnamefont {D.}~\bibnamefont
  {Campagnari}}, \ and\ \bibinfo {author} {\bibfnamefont {A.}~\bibnamefont
  {Szczepaniak}},\ }\href {\doibase 10.1103/PhysRevD.84.045006} {\bibfield
  {journal} {\bibinfo  {journal} {Phys.Rev.}\ }\textbf {\bibinfo {volume}
  {D84}},\ \bibinfo {pages} {045006} (\bibinfo {year} {2011})},\ \Eprint
  {http://arxiv.org/abs/1107.3389} {arXiv:1107.3389 [hep-th]} \BibitemShut
  {NoStop}%
\bibitem [{\citenamefont {Heffner}\ \emph {et~al.}(2012)\citenamefont
  {Heffner}, \citenamefont {Reinhardt},\ and\ \citenamefont
  {Campagnari}}]{Heffner:2012sx}%
  \BibitemOpen
  \bibfield  {author} {\bibinfo {author} {\bibfnamefont {J.}~\bibnamefont
  {Heffner}}, \bibinfo {author} {\bibfnamefont {H.}~\bibnamefont {Reinhardt}},
  \ and\ \bibinfo {author} {\bibfnamefont {D.~R.}\ \bibnamefont {Campagnari}},\
  }\href {\doibase 10.1103/PhysRevD.85.125029} {\bibfield  {journal} {\bibinfo
  {journal} {Phys.Rev.}\ }\textbf {\bibinfo {volume} {D85}},\ \bibinfo {pages}
  {125029} (\bibinfo {year} {2012})},\ \Eprint {http://arxiv.org/abs/1206.3936}
  {arXiv:1206.3936 [hep-th]} \BibitemShut {NoStop}%
\bibitem [{\citenamefont {Reinhardt}\ and\ \citenamefont
  {Heffner}(2012)}]{Reinhardt:2012qe}%
  \BibitemOpen
  \bibfield  {author} {\bibinfo {author} {\bibfnamefont {H.}~\bibnamefont
  {Reinhardt}}\ and\ \bibinfo {author} {\bibfnamefont {J.}~\bibnamefont
  {Heffner}},\ }\href {\doibase 10.1016/j.physletb.2012.10.084} {\bibfield
  {journal} {\bibinfo  {journal} {Phys.Lett.}\ }\textbf {\bibinfo {volume}
  {B718}},\ \bibinfo {pages} {672} (\bibinfo {year} {2012})},\ \Eprint
  {http://arxiv.org/abs/1210.1742} {arXiv:1210.1742 [hep-th]} \BibitemShut
  {NoStop}%
\bibitem [{\citenamefont {Reinhardt}\ and\ \citenamefont
  {Heffner}(2013)}]{Reinhardt:2013iia}%
  \BibitemOpen
  \bibfield  {author} {\bibinfo {author} {\bibfnamefont {H.}~\bibnamefont
  {Reinhardt}}\ and\ \bibinfo {author} {\bibfnamefont {J.}~\bibnamefont
  {Heffner}},\ }\href {\doibase 10.1103/PhysRevD.88.045024} {\bibfield
  {journal} {\bibinfo  {journal} {Phys.Rev.}\ }\textbf {\bibinfo {volume}
  {D88}},\ \bibinfo {pages} {045024} (\bibinfo {year} {2013})},\ \Eprint
  {http://arxiv.org/abs/1304.2980} {arXiv:1304.2980 [hep-th]} \BibitemShut
  {NoStop}%
\bibitem [{\citenamefont {Burgio}\ \emph {et~al.}(2000)\citenamefont {Burgio},
  \citenamefont {De~Pietri}, \citenamefont {Morales-Tecotl}, \citenamefont
  {Urrutia},\ and\ \citenamefont {Vergara}}]{Burgio:1999tg}%
  \BibitemOpen
  \bibfield  {author} {\bibinfo {author} {\bibfnamefont {G.}~\bibnamefont
  {Burgio}}, \bibinfo {author} {\bibfnamefont {R.}~\bibnamefont {De~Pietri}},
  \bibinfo {author} {\bibfnamefont {H.~A.}\ \bibnamefont {Morales-Tecotl}},
  \bibinfo {author} {\bibfnamefont {L.~F.}\ \bibnamefont {Urrutia}}, \ and\
  \bibinfo {author} {\bibfnamefont {J.~D.}\ \bibnamefont {Vergara}},\ }\href
  {\doibase 10.1016/S0550-3213(99)00533-7} {\bibfield  {journal} {\bibinfo
  {journal} {Nucl. Phys.}\ }\textbf {\bibinfo {volume} {B566}},\ \bibinfo
  {pages} {547} (\bibinfo {year} {2000})},\ \Eprint
  {http://arxiv.org/abs/hep-lat/9906036} {arXiv:hep-lat/9906036} \BibitemShut
  {NoStop}%
\bibitem [{\citenamefont {Gribov}(1978)}]{Gribov:1977wm}%
  \BibitemOpen
  \bibfield  {author} {\bibinfo {author} {\bibfnamefont {V.~N.}\ \bibnamefont
  {Gribov}},\ }\href {\doibase 10.1016/0550-3213(78)90175-X} {\bibfield
  {journal} {\bibinfo  {journal} {Nucl. Phys.}\ }\textbf {\bibinfo {volume}
  {B139}},\ \bibinfo {pages} {1} (\bibinfo {year} {1978})}\BibitemShut
  {NoStop}%
\bibitem [{\citenamefont {Wilson}(1974)}]{Wilson:1974sk}%
  \BibitemOpen
  \bibfield  {author} {\bibinfo {author} {\bibfnamefont {K.~G.}\ \bibnamefont
  {Wilson}},\ }\href {\doibase 10.1103/PhysRevD.10.2445} {\bibfield  {journal}
  {\bibinfo  {journal} {Phys.Rev.}\ }\textbf {\bibinfo {volume} {D10}},\
  \bibinfo {pages} {2445} (\bibinfo {year} {1974})}\BibitemShut {NoStop}%
\bibitem [{\citenamefont {Cornwall}(1979)}]{Cornwall:1979hz}%
  \BibitemOpen
  \bibfield  {author} {\bibinfo {author} {\bibfnamefont {J.~M.}\ \bibnamefont
  {Cornwall}},\ }\href {\doibase 10.1016/0550-3213(79)90111-1} {\bibfield
  {journal} {\bibinfo  {journal} {Nucl. Phys.}\ }\textbf {\bibinfo {volume}
  {B157}},\ \bibinfo {pages} {392} (\bibinfo {year} {1979})}\BibitemShut
  {NoStop}%
\bibitem [{\citenamefont {Zwanziger}(1994)}]{Zwanziger:1993dh}%
  \BibitemOpen
  \bibfield  {author} {\bibinfo {author} {\bibfnamefont {D.}~\bibnamefont
  {Zwanziger}},\ }\href {\doibase 10.1016/0550-3213(94)90396-4} {\bibfield
  {journal} {\bibinfo  {journal} {Nucl.Phys.}\ }\textbf {\bibinfo {volume}
  {B412}},\ \bibinfo {pages} {657} (\bibinfo {year} {1994})}\BibitemShut
  {NoStop}%
\bibitem [{\citenamefont {Zwanziger}(1997)}]{Zwanziger:1995cv}%
  \BibitemOpen
  \bibfield  {author} {\bibinfo {author} {\bibfnamefont {D.}~\bibnamefont
  {Zwanziger}},\ }\href {\doibase 10.1016/S0550-3213(96)00566-4} {\bibfield
  {journal} {\bibinfo  {journal} {Nucl. Phys.}\ }\textbf {\bibinfo {volume}
  {B485}},\ \bibinfo {pages} {185} (\bibinfo {year} {1997})},\ \Eprint
  {http://arxiv.org/abs/hep-th/9603203} {arXiv:hep-th/9603203} \BibitemShut
  {NoStop}%
\bibitem [{\citenamefont {Cucchieri}\ and\ \citenamefont
  {Zwanziger}(2002{\natexlab{a}})}]{Cucchieri:2000gu}%
  \BibitemOpen
  \bibfield  {author} {\bibinfo {author} {\bibfnamefont {A.}~\bibnamefont
  {Cucchieri}}\ and\ \bibinfo {author} {\bibfnamefont {D.}~\bibnamefont
  {Zwanziger}},\ }\href {\doibase 10.1103/PhysRevD.65.014001} {\bibfield
  {journal} {\bibinfo  {journal} {Phys. Rev.}\ }\textbf {\bibinfo {volume}
  {D65}},\ \bibinfo {pages} {014001} (\bibinfo {year} {2002}{\natexlab{a}})},\
  \Eprint {http://arxiv.org/abs/hep-lat/0008026} {arXiv:hep-lat/0008026}
  \BibitemShut {NoStop}%
\bibitem [{\citenamefont {Cucchieri}\ and\ \citenamefont
  {Zwanziger}(2002{\natexlab{b}})}]{Cucchieri:2000kw}%
  \BibitemOpen
  \bibfield  {author} {\bibinfo {author} {\bibfnamefont {A.}~\bibnamefont
  {Cucchieri}}\ and\ \bibinfo {author} {\bibfnamefont {D.}~\bibnamefont
  {Zwanziger}},\ }\href {\doibase 10.1016/S0370-2693(01)01365-X} {\bibfield
  {journal} {\bibinfo  {journal} {Phys. Lett.}\ }\textbf {\bibinfo {volume}
  {B524}},\ \bibinfo {pages} {123} (\bibinfo {year} {2002}{\natexlab{b}})},\
  \Eprint {http://arxiv.org/abs/hep-lat/0012024} {arXiv:hep-lat/0012024}
  \BibitemShut {NoStop}%
\bibitem [{\citenamefont {Greensite}\ and\ \citenamefont
  {Olejnik}(2003)}]{Greensite:2003xf}%
  \BibitemOpen
  \bibfield  {author} {\bibinfo {author} {\bibfnamefont {J.}~\bibnamefont
  {Greensite}}\ and\ \bibinfo {author} {\bibfnamefont {S.}~\bibnamefont
  {Olejnik}},\ }\href {\doibase 10.1103/PhysRevD.67.094503} {\bibfield
  {journal} {\bibinfo  {journal} {Phys.Rev.}\ }\textbf {\bibinfo {volume}
  {D67}},\ \bibinfo {pages} {094503} (\bibinfo {year} {2003})},\ \Eprint
  {http://arxiv.org/abs/hep-lat/0302018} {arXiv:hep-lat/0302018 [hep-lat]}
  \BibitemShut {NoStop}%
\bibitem [{\citenamefont {Langfeld}\ and\ \citenamefont
  {Moyaerts}(2004)}]{Langfeld:2004qs}%
  \BibitemOpen
  \bibfield  {author} {\bibinfo {author} {\bibfnamefont {K.}~\bibnamefont
  {Langfeld}}\ and\ \bibinfo {author} {\bibfnamefont {L.}~\bibnamefont
  {Moyaerts}},\ }\href {\doibase 10.1103/PhysRevD.70.074507} {\bibfield
  {journal} {\bibinfo  {journal} {Phys. Rev.}\ }\textbf {\bibinfo {volume}
  {D70}},\ \bibinfo {pages} {074507} (\bibinfo {year} {2004})},\ \Eprint
  {http://arxiv.org/abs/hep-lat/0406024} {arXiv:hep-lat/0406024} \BibitemShut
  {NoStop}%
\bibitem [{\citenamefont {Greensite}\ \emph {et~al.}(2004)\citenamefont
  {Greensite}, \citenamefont {Olejnik},\ and\ \citenamefont
  {Zwanziger}}]{Greensite:2004ke}%
  \BibitemOpen
  \bibfield  {author} {\bibinfo {author} {\bibfnamefont {J.}~\bibnamefont
  {Greensite}}, \bibinfo {author} {\bibfnamefont {S.}~\bibnamefont {Olejnik}},
  \ and\ \bibinfo {author} {\bibfnamefont {D.}~\bibnamefont {Zwanziger}},\
  }\href {\doibase 10.1103/PhysRevD.69.074506} {\bibfield  {journal} {\bibinfo
  {journal} {Phys.Rev.}\ }\textbf {\bibinfo {volume} {D69}},\ \bibinfo {pages}
  {074506} (\bibinfo {year} {2004})},\ \Eprint
  {http://arxiv.org/abs/hep-lat/0401003} {arXiv:hep-lat/0401003 [hep-lat]}
  \BibitemShut {NoStop}%
\bibitem [{\citenamefont {Burgio}\ \emph {et~al.}(2009)\citenamefont {Burgio},
  \citenamefont {Quandt},\ and\ \citenamefont {Reinhardt}}]{Burgio:2008jr}%
  \BibitemOpen
  \bibfield  {author} {\bibinfo {author} {\bibfnamefont {G.}~\bibnamefont
  {Burgio}}, \bibinfo {author} {\bibfnamefont {M.}~\bibnamefont {Quandt}}, \
  and\ \bibinfo {author} {\bibfnamefont {H.}~\bibnamefont {Reinhardt}},\ }\href
  {\doibase 10.1103/PhysRevLett.102.032002} {\bibfield  {journal} {\bibinfo
  {journal} {Phys. Rev. Lett.}\ }\textbf {\bibinfo {volume} {102}},\ \bibinfo
  {pages} {032002} (\bibinfo {year} {2009})},\ \Eprint
  {http://arxiv.org/abs/0807.3291} {arXiv:0807.3291 [hep-lat]} \BibitemShut
  {NoStop}%
\bibitem [{\citenamefont {Voigt}\ \emph {et~al.}(2008)\citenamefont {Voigt},
  \citenamefont {Ilgenfritz}, \citenamefont {Muller-Preussker},\ and\
  \citenamefont {Sternbeck}}]{Voigt:2008rr}%
  \BibitemOpen
  \bibfield  {author} {\bibinfo {author} {\bibfnamefont {A.}~\bibnamefont
  {Voigt}}, \bibinfo {author} {\bibfnamefont {E.-M.}\ \bibnamefont
  {Ilgenfritz}}, \bibinfo {author} {\bibfnamefont {M.}~\bibnamefont
  {Muller-Preussker}}, \ and\ \bibinfo {author} {\bibfnamefont
  {A.}~\bibnamefont {Sternbeck}},\ }\href {\doibase 10.1103/PhysRevD.78.014501}
  {\bibfield  {journal} {\bibinfo  {journal} {Phys.Rev.}\ }\textbf {\bibinfo
  {volume} {D78}},\ \bibinfo {pages} {014501} (\bibinfo {year} {2008})},\
  \Eprint {http://arxiv.org/abs/0803.2307} {arXiv:0803.2307 [hep-lat]}
  \BibitemShut {NoStop}%
\bibitem [{\citenamefont {Greensite}(2009)}]{Greensite:2009eb}%
  \BibitemOpen
  \bibfield  {author} {\bibinfo {author} {\bibfnamefont {J.}~\bibnamefont
  {Greensite}},\ }\href {\doibase 10.1103/PhysRevD.80.045003} {\bibfield
  {journal} {\bibinfo  {journal} {Phys.Rev.}\ }\textbf {\bibinfo {volume}
  {D80}},\ \bibinfo {pages} {045003} (\bibinfo {year} {2009})},\ \Eprint
  {http://arxiv.org/abs/0903.2501} {arXiv:0903.2501 [hep-lat]} \BibitemShut
  {NoStop}%
\bibitem [{\citenamefont {Burgio}\ \emph {et~al.}(2010)\citenamefont {Burgio},
  \citenamefont {Quandt},\ and\ \citenamefont {Reinhardt}}]{Burgio:2009xp}%
  \BibitemOpen
  \bibfield  {author} {\bibinfo {author} {\bibfnamefont {G.}~\bibnamefont
  {Burgio}}, \bibinfo {author} {\bibfnamefont {M.}~\bibnamefont {Quandt}}, \
  and\ \bibinfo {author} {\bibfnamefont {H.}~\bibnamefont {Reinhardt}},\ }\href
  {\doibase 10.1103/PhysRevD.81.074502} {\bibfield  {journal} {\bibinfo
  {journal} {Phys.Rev.}\ }\textbf {\bibinfo {volume} {D81}},\ \bibinfo {pages}
  {074502} (\bibinfo {year} {2010})},\ \Eprint {http://arxiv.org/abs/0911.5101}
  {arXiv:0911.5101 [hep-lat]} \BibitemShut {NoStop}%
\bibitem [{\citenamefont {Nakagawa}\ \emph {et~al.}(2008)\citenamefont
  {Nakagawa}, \citenamefont {Nakamura}, \citenamefont {Saito},\ and\
  \citenamefont {Toki}}]{Nakagawa:2008zza}%
  \BibitemOpen
  \bibfield  {author} {\bibinfo {author} {\bibfnamefont {Y.}~\bibnamefont
  {Nakagawa}}, \bibinfo {author} {\bibfnamefont {A.}~\bibnamefont {Nakamura}},
  \bibinfo {author} {\bibfnamefont {T.}~\bibnamefont {Saito}}, \ and\ \bibinfo
  {author} {\bibfnamefont {T.}~\bibnamefont {Toki}},\ }\href {\doibase
  10.1142/S0217732308029356} {\bibfield  {journal} {\bibinfo  {journal}
  {Mod.Phys.Lett.}\ }\textbf {\bibinfo {volume} {A23}},\ \bibinfo {pages}
  {2348} (\bibinfo {year} {2008})}\BibitemShut {NoStop}%
\bibitem [{\citenamefont {Nakagawa}\ \emph {et~al.}(2009)\citenamefont
  {Nakagawa}, \citenamefont {Nakamura}, \citenamefont {Saito},\ and\
  \citenamefont {Toki}}]{Nakagawa:2009is}%
  \BibitemOpen
  \bibfield  {author} {\bibinfo {author} {\bibfnamefont {Y.}~\bibnamefont
  {Nakagawa}}, \bibinfo {author} {\bibfnamefont {A.}~\bibnamefont {Nakamura}},
  \bibinfo {author} {\bibfnamefont {T.}~\bibnamefont {Saito}}, \ and\ \bibinfo
  {author} {\bibfnamefont {H.}~\bibnamefont {Toki}},\ }\href@noop {} {\bibfield
   {journal} {\bibinfo  {journal} {PoS}\ }\textbf {\bibinfo {volume}
  {LAT2009}},\ \bibinfo {pages} {230} (\bibinfo {year} {2009})},\ \Eprint
  {http://arxiv.org/abs/0911.2550} {arXiv:0911.2550 [hep-lat]} \BibitemShut
  {NoStop}%
\bibitem [{\citenamefont {Quandt}\ \emph {et~al.}(2010)\citenamefont {Quandt},
  \citenamefont {Reinhardt},\ and\ \citenamefont {Burgio}}]{Quandt:2010yq}%
  \BibitemOpen
  \bibfield  {author} {\bibinfo {author} {\bibfnamefont {M.}~\bibnamefont
  {Quandt}}, \bibinfo {author} {\bibfnamefont {H.}~\bibnamefont {Reinhardt}}, \
  and\ \bibinfo {author} {\bibfnamefont {G.}~\bibnamefont {Burgio}},\ }\href
  {\doibase 10.1103/PhysRevD.81.065016} {\bibfield  {journal} {\bibinfo
  {journal} {Phys.Rev.}\ }\textbf {\bibinfo {volume} {D81}},\ \bibinfo {pages}
  {065016} (\bibinfo {year} {2010})},\ \Eprint {http://arxiv.org/abs/1001.3699}
  {arXiv:1001.3699 [hep-lat]} \BibitemShut {NoStop}%
\bibitem [{\citenamefont {Nakagawa}\ \emph {et~al.}(2010)\citenamefont
  {Nakagawa}, \citenamefont {Nakamura}, \citenamefont {Saito},\ and\
  \citenamefont {Toki}}]{Nakagawa:2010eh}%
  \BibitemOpen
  \bibfield  {author} {\bibinfo {author} {\bibfnamefont {Y.}~\bibnamefont
  {Nakagawa}}, \bibinfo {author} {\bibfnamefont {A.}~\bibnamefont {Nakamura}},
  \bibinfo {author} {\bibfnamefont {T.}~\bibnamefont {Saito}}, \ and\ \bibinfo
  {author} {\bibfnamefont {H.}~\bibnamefont {Toki}},\ }\href {\doibase
  10.1103/PhysRevD.81.054509} {\bibfield  {journal} {\bibinfo  {journal}
  {Phys.Rev.}\ }\textbf {\bibinfo {volume} {D81}},\ \bibinfo {pages} {054509}
  (\bibinfo {year} {2010})},\ \Eprint {http://arxiv.org/abs/1003.4792}
  {arXiv:1003.4792 [hep-lat]} \BibitemShut {NoStop}%
\bibitem [{\citenamefont {Iritani}\ and\ \citenamefont
  {Suganuma}(2011)}]{Iritani:2010mu}%
  \BibitemOpen
  \bibfield  {author} {\bibinfo {author} {\bibfnamefont {T.}~\bibnamefont
  {Iritani}}\ and\ \bibinfo {author} {\bibfnamefont {H.}~\bibnamefont
  {Suganuma}},\ }\href {\doibase 10.1103/PhysRevD.83.054502} {\bibfield
  {journal} {\bibinfo  {journal} {Phys.Rev.}\ }\textbf {\bibinfo {volume}
  {D83}},\ \bibinfo {pages} {054502} (\bibinfo {year} {2011})},\ \Eprint
  {http://arxiv.org/abs/1011.4767} {arXiv:1011.4767 [hep-lat]} \BibitemShut
  {NoStop}%
\bibitem [{\citenamefont {Reinhardt}\ \emph {et~al.}(2012)\citenamefont
  {Reinhardt}, \citenamefont {Quandt},\ and\ \citenamefont
  {Burgio}}]{Reinhardt:2011fq}%
  \BibitemOpen
  \bibfield  {author} {\bibinfo {author} {\bibfnamefont {H.}~\bibnamefont
  {Reinhardt}}, \bibinfo {author} {\bibfnamefont {M.}~\bibnamefont {Quandt}}, \
  and\ \bibinfo {author} {\bibfnamefont {G.}~\bibnamefont {Burgio}},\ }\href
  {\doibase 10.1103/PhysRevD.85.025001} {\bibfield  {journal} {\bibinfo
  {journal} {Phys. Rev. D}\ }\textbf {\bibinfo {volume} {85}},\ \bibinfo
  {pages} {025001} (\bibinfo {year} {2012})},\ \Eprint
  {http://arxiv.org/abs/1110.2927} {arXiv:1110.2927 [hep-th]} \BibitemShut
  {NoStop}%
\bibitem [{\citenamefont {Nakagawa}\ \emph {et~al.}(2011)\citenamefont
  {Nakagawa}, \citenamefont {Nakamura}, \citenamefont {Saito},\ and\
  \citenamefont {Toki}}]{Nakagawa:2011ar}%
  \BibitemOpen
  \bibfield  {author} {\bibinfo {author} {\bibfnamefont {Y.}~\bibnamefont
  {Nakagawa}}, \bibinfo {author} {\bibfnamefont {A.}~\bibnamefont {Nakamura}},
  \bibinfo {author} {\bibfnamefont {T.}~\bibnamefont {Saito}}, \ and\ \bibinfo
  {author} {\bibfnamefont {H.}~\bibnamefont {Toki}},\ }\href {\doibase
  10.1103/PhysRevD.83.114503} {\bibfield  {journal} {\bibinfo  {journal}
  {Phys.Rev.}\ }\textbf {\bibinfo {volume} {D83}},\ \bibinfo {pages} {114503}
  (\bibinfo {year} {2011})},\ \Eprint {http://arxiv.org/abs/1105.6185}
  {arXiv:1105.6185 [hep-lat]} \BibitemShut {NoStop}%
\bibitem [{\citenamefont {Burgio}\ \emph
  {et~al.}(2012{\natexlab{a}})\citenamefont {Burgio}, \citenamefont {Quandt},\
  and\ \citenamefont {Reinhardt}}]{Burgio:2012bk}%
  \BibitemOpen
  \bibfield  {author} {\bibinfo {author} {\bibfnamefont {G.}~\bibnamefont
  {Burgio}}, \bibinfo {author} {\bibfnamefont {M.}~\bibnamefont {Quandt}}, \
  and\ \bibinfo {author} {\bibfnamefont {H.}~\bibnamefont {Reinhardt}},\ }\href
  {\doibase 10.1103/PhysRevD.86.045029} {\bibfield  {journal} {\bibinfo
  {journal} {Phys.Rev.}\ }\textbf {\bibinfo {volume} {D86}},\ \bibinfo {pages}
  {045029} (\bibinfo {year} {2012}{\natexlab{a}})},\ \Eprint
  {http://arxiv.org/abs/1205.5674} {arXiv:1205.5674 [hep-lat]} \BibitemShut
  {NoStop}%
\bibitem [{\citenamefont {Burgio}\ \emph
  {et~al.}(2012{\natexlab{b}})\citenamefont {Burgio}, \citenamefont {Schrock},
  \citenamefont {Reinhardt},\ and\ \citenamefont {Quandt}}]{Burgio:2012ph}%
  \BibitemOpen
  \bibfield  {author} {\bibinfo {author} {\bibfnamefont {G.}~\bibnamefont
  {Burgio}}, \bibinfo {author} {\bibfnamefont {M.}~\bibnamefont {Schrock}},
  \bibinfo {author} {\bibfnamefont {H.}~\bibnamefont {Reinhardt}}, \ and\
  \bibinfo {author} {\bibfnamefont {M.}~\bibnamefont {Quandt}},\ }\href
  {\doibase 10.1103/PhysRevD.86.014506} {\bibfield  {journal} {\bibinfo
  {journal} {Phys.Rev.}\ }\textbf {\bibinfo {volume} {D86}},\ \bibinfo {pages}
  {014506} (\bibinfo {year} {2012}{\natexlab{b}})},\ \Eprint
  {http://arxiv.org/abs/1204.0716} {arXiv:1204.0716 [hep-lat]} \BibitemShut
  {NoStop}%
\bibitem [{\citenamefont {Iritani}\ and\ \citenamefont
  {Suganuma}(2012)}]{Iritani:2012bc}%
  \BibitemOpen
  \bibfield  {author} {\bibinfo {author} {\bibfnamefont {T.}~\bibnamefont
  {Iritani}}\ and\ \bibinfo {author} {\bibfnamefont {H.}~\bibnamefont
  {Suganuma}},\ }\href {\doibase 10.1103/PhysRevD.86.074034} {\bibfield
  {journal} {\bibinfo  {journal} {Phys.Rev.}\ }\textbf {\bibinfo {volume}
  {D86}},\ \bibinfo {pages} {074034} (\bibinfo {year} {2012})},\ \Eprint
  {http://arxiv.org/abs/1204.6591} {arXiv:1204.6591 [hep-lat]} \BibitemShut
  {NoStop}%
\bibitem [{\citenamefont {Burgio}\ and\ \citenamefont
  {al.}(2003)}]{Burgio:2003in}%
  \BibitemOpen
  \bibfield  {author} {\bibinfo {author} {\bibfnamefont {G.}~\bibnamefont
  {Burgio}}\ and\ \bibinfo {author} {\bibnamefont {al.}} (\bibinfo
  {collaboration} {TrinLat}),\ }\href {\doibase 10.1103/PhysRevD.67.114502}
  {\bibfield  {journal} {\bibinfo  {journal} {Phys. Rev.}\ }\textbf {\bibinfo
  {volume} {D67}},\ \bibinfo {pages} {114502} (\bibinfo {year} {2003})},\
  \Eprint {http://arxiv.org/abs/hep-lat/0303005} {arXiv:hep-lat/0303005}
  \BibitemShut {NoStop}%
\bibitem [{\citenamefont {Luscher}\ and\ \citenamefont
  {Weisz}(2001)}]{Luscher:2001up}%
  \BibitemOpen
  \bibfield  {author} {\bibinfo {author} {\bibfnamefont {M.}~\bibnamefont
  {Luscher}}\ and\ \bibinfo {author} {\bibfnamefont {P.}~\bibnamefont
  {Weisz}},\ }\href {\doibase 10.1088/1126-6708/2001/09/010} {\bibfield
  {journal} {\bibinfo  {journal} {JHEP}\ }\textbf {\bibinfo {volume} {0109}},\
  \bibinfo {pages} {010} (\bibinfo {year} {2001})},\ \Eprint
  {http://arxiv.org/abs/hep-lat/0108014} {arXiv:hep-lat/0108014 [hep-lat]}
  \BibitemShut {NoStop}%
\bibitem [{\citenamefont {Vogt}\ \emph {et~al.}(2014)\citenamefont {Vogt},
  \citenamefont {Burgio}, \citenamefont {Quandt},\ and\ \citenamefont
  {Reinhardt}}]{Vogt:2013jha}%
  \BibitemOpen
  \bibfield  {author} {\bibinfo {author} {\bibfnamefont {H.}~\bibnamefont
  {Vogt}}, \bibinfo {author} {\bibfnamefont {G.}~\bibnamefont {Burgio}},
  \bibinfo {author} {\bibfnamefont {M.}~\bibnamefont {Quandt}}, \ and\ \bibinfo
  {author} {\bibfnamefont {H.}~\bibnamefont {Reinhardt}},\ }\href@noop {}
  {\bibfield  {journal} {\bibinfo  {journal} {PoS}\ }\textbf {\bibinfo {volume}
  {LATTICE2013}},\ \bibinfo {pages} {363} (\bibinfo {year} {2014})},\ \Eprint
  {http://arxiv.org/abs/1311.5707} {arXiv:1311.5707 [hep-lat]} \BibitemShut
  {NoStop}%
\bibitem [{\citenamefont {de~Forcrand}\ and\ \citenamefont
  {D'Elia}(1999)}]{deForcrand:1999ms}%
  \BibitemOpen
  \bibfield  {author} {\bibinfo {author} {\bibfnamefont {P.}~\bibnamefont
  {de~Forcrand}}\ and\ \bibinfo {author} {\bibfnamefont {M.}~\bibnamefont
  {D'Elia}},\ }\href {\doibase 10.1103/PhysRevLett.82.4582} {\bibfield
  {journal} {\bibinfo  {journal} {Phys.Rev.Lett.}\ }\textbf {\bibinfo {volume}
  {82}},\ \bibinfo {pages} {4582} (\bibinfo {year} {1999})},\ \Eprint
  {http://arxiv.org/abs/hep-lat/9901020} {arXiv:hep-lat/9901020 [hep-lat]}
  \BibitemShut {NoStop}%
\bibitem [{\citenamefont {Del~Debbio}\ \emph {et~al.}(1997)\citenamefont
  {Del~Debbio}, \citenamefont {Faber}, \citenamefont {Greensite},\ and\
  \citenamefont {Olejnik}}]{DelDebbio:1996mh}%
  \BibitemOpen
  \bibfield  {author} {\bibinfo {author} {\bibfnamefont {L.}~\bibnamefont
  {Del~Debbio}}, \bibinfo {author} {\bibfnamefont {M.}~\bibnamefont {Faber}},
  \bibinfo {author} {\bibfnamefont {J.}~\bibnamefont {Greensite}}, \ and\
  \bibinfo {author} {\bibfnamefont {S.}~\bibnamefont {Olejnik}},\ }\href
  {\doibase 10.1103/PhysRevD.55.2298} {\bibfield  {journal} {\bibinfo
  {journal} {Phys.Rev.}\ }\textbf {\bibinfo {volume} {D55}},\ \bibinfo {pages}
  {2298} (\bibinfo {year} {1997})},\ \Eprint
  {http://arxiv.org/abs/hep-lat/9610005} {arXiv:hep-lat/9610005 [hep-lat]}
  \BibitemShut {NoStop}%
\bibitem [{\citenamefont {Langfeld}\ \emph {et~al.}(1998)\citenamefont
  {Langfeld}, \citenamefont {Reinhardt},\ and\ \citenamefont
  {Tennert}}]{Langfeld:1997jx}%
  \BibitemOpen
  \bibfield  {author} {\bibinfo {author} {\bibfnamefont {K.}~\bibnamefont
  {Langfeld}}, \bibinfo {author} {\bibfnamefont {H.}~\bibnamefont {Reinhardt}},
  \ and\ \bibinfo {author} {\bibfnamefont {O.}~\bibnamefont {Tennert}},\ }\href
  {\doibase 10.1016/S0370-2693(97)01435-4} {\bibfield  {journal} {\bibinfo
  {journal} {Phys.Lett.}\ }\textbf {\bibinfo {volume} {B419}},\ \bibinfo
  {pages} {317} (\bibinfo {year} {1998})},\ \Eprint
  {http://arxiv.org/abs/hep-lat/9710068} {arXiv:hep-lat/9710068 [hep-lat]}
  \BibitemShut {NoStop}%
\bibitem [{\citenamefont {Del~Debbio}\ \emph {et~al.}(1998)\citenamefont
  {Del~Debbio}, \citenamefont {Faber}, \citenamefont {Giedt}, \citenamefont
  {Greensite},\ and\ \citenamefont {Olejnik}}]{DelDebbio:1998uu}%
  \BibitemOpen
  \bibfield  {author} {\bibinfo {author} {\bibfnamefont {L.}~\bibnamefont
  {Del~Debbio}}, \bibinfo {author} {\bibfnamefont {M.}~\bibnamefont {Faber}},
  \bibinfo {author} {\bibfnamefont {J.}~\bibnamefont {Giedt}}, \bibinfo
  {author} {\bibfnamefont {J.}~\bibnamefont {Greensite}}, \ and\ \bibinfo
  {author} {\bibfnamefont {S.}~\bibnamefont {Olejnik}},\ }\href {\doibase
  10.1103/PhysRevD.58.094501} {\bibfield  {journal} {\bibinfo  {journal}
  {Phys.Rev.}\ }\textbf {\bibinfo {volume} {D58}},\ \bibinfo {pages} {094501}
  (\bibinfo {year} {1998})},\ \Eprint {http://arxiv.org/abs/hep-lat/9801027}
  {arXiv:hep-lat/9801027 [hep-lat]} \BibitemShut {NoStop}%
\bibitem [{\citenamefont {Engelhardt}\ \emph {et~al.}(2000)\citenamefont
  {Engelhardt}, \citenamefont {Langfeld}, \citenamefont {Reinhardt},\ and\
  \citenamefont {Tennert}}]{Engelhardt:1999fd}%
  \BibitemOpen
  \bibfield  {author} {\bibinfo {author} {\bibfnamefont {M.}~\bibnamefont
  {Engelhardt}}, \bibinfo {author} {\bibfnamefont {K.}~\bibnamefont
  {Langfeld}}, \bibinfo {author} {\bibfnamefont {H.}~\bibnamefont {Reinhardt}},
  \ and\ \bibinfo {author} {\bibfnamefont {O.}~\bibnamefont {Tennert}},\ }\href
  {\doibase 10.1103/PhysRevD.61.054504} {\bibfield  {journal} {\bibinfo
  {journal} {Phys.Rev.}\ }\textbf {\bibinfo {volume} {D61}},\ \bibinfo {pages}
  {054504} (\bibinfo {year} {2000})},\ \Eprint
  {http://arxiv.org/abs/hep-lat/9904004} {arXiv:hep-lat/9904004 [hep-lat]}
  \BibitemShut {NoStop}%
\bibitem [{\citenamefont {'t~Hooft}(1979)}]{'tHooft:1979uj}%
  \BibitemOpen
  \bibfield  {author} {\bibinfo {author} {\bibfnamefont {G.}~\bibnamefont
  {'t~Hooft}},\ }\href {\doibase 10.1016/0550-3213(79)90595-9} {\bibfield
  {journal} {\bibinfo  {journal} {Nucl.Phys.}\ }\textbf {\bibinfo {volume}
  {B153}},\ \bibinfo {pages} {141} (\bibinfo {year} {1979})}\BibitemShut
  {NoStop}%
\bibitem [{\citenamefont {Hart}\ \emph {et~al.}(2000)\citenamefont {Hart},
  \citenamefont {Lucini}, \citenamefont {Schram},\ and\ \citenamefont
  {Teper}}]{Hart:2000en}%
  \BibitemOpen
  \bibfield  {author} {\bibinfo {author} {\bibfnamefont {A.}~\bibnamefont
  {Hart}}, \bibinfo {author} {\bibfnamefont {B.}~\bibnamefont {Lucini}},
  \bibinfo {author} {\bibfnamefont {Z.}~\bibnamefont {Schram}}, \ and\ \bibinfo
  {author} {\bibfnamefont {M.}~\bibnamefont {Teper}},\ }\href {\doibase
  10.1088/1126-6708/2000/06/040} {\bibfield  {journal} {\bibinfo  {journal}
  {JHEP}\ }\textbf {\bibinfo {volume} {0006}},\ \bibinfo {pages} {040}
  (\bibinfo {year} {2000})},\ \Eprint {http://arxiv.org/abs/hep-lat/0005010}
  {arXiv:hep-lat/0005010 [hep-lat]} \BibitemShut {NoStop}%
\bibitem [{\citenamefont {Kovacs}\ and\ \citenamefont
  {Tomboulis}(2000)}]{Kovacs:2000sy}%
  \BibitemOpen
  \bibfield  {author} {\bibinfo {author} {\bibfnamefont {T.~G.}\ \bibnamefont
  {Kovacs}}\ and\ \bibinfo {author} {\bibfnamefont {E.}~\bibnamefont
  {Tomboulis}},\ }\href {\doibase 10.1103/PhysRevLett.85.704} {\bibfield
  {journal} {\bibinfo  {journal} {Phys.Rev.Lett.}\ }\textbf {\bibinfo {volume}
  {85}},\ \bibinfo {pages} {704} (\bibinfo {year} {2000})},\ \Eprint
  {http://arxiv.org/abs/hep-lat/0002004} {arXiv:hep-lat/0002004 [hep-lat]}
  \BibitemShut {NoStop}%
\bibitem [{\citenamefont {de~Forcrand}\ \emph {et~al.}(2001)\citenamefont
  {de~Forcrand}, \citenamefont {D'Elia},\ and\ \citenamefont
  {Pepe}}]{deForcrand:2000fi}%
  \BibitemOpen
  \bibfield  {author} {\bibinfo {author} {\bibfnamefont {P.}~\bibnamefont
  {de~Forcrand}}, \bibinfo {author} {\bibfnamefont {M.}~\bibnamefont {D'Elia}},
  \ and\ \bibinfo {author} {\bibfnamefont {M.}~\bibnamefont {Pepe}},\ }\href
  {\doibase 10.1103/PhysRevLett.86.1438} {\bibfield  {journal} {\bibinfo
  {journal} {Phys.Rev.Lett.}\ }\textbf {\bibinfo {volume} {86}},\ \bibinfo
  {pages} {1438} (\bibinfo {year} {2001})},\ \Eprint
  {http://arxiv.org/abs/hep-lat/0007034} {arXiv:hep-lat/0007034 [hep-lat]}
  \BibitemShut {NoStop}%
\bibitem [{\citenamefont {Barresi}\ \emph {et~al.}(2002)\citenamefont
  {Barresi}, \citenamefont {Burgio},\ and\ \citenamefont
  {Muller-Preussker}}]{Barresi:2001dt}%
  \BibitemOpen
  \bibfield  {author} {\bibinfo {author} {\bibfnamefont {A.}~\bibnamefont
  {Barresi}}, \bibinfo {author} {\bibfnamefont {G.}~\bibnamefont {Burgio}}, \
  and\ \bibinfo {author} {\bibfnamefont {M.}~\bibnamefont {Muller-Preussker}},\
  }\href {\doibase 10.1016/S0920-5632(01)01758-3} {\bibfield  {journal}
  {\bibinfo  {journal} {Nucl.Phys.Proc.Suppl.}\ }\textbf {\bibinfo {volume}
  {106}},\ \bibinfo {pages} {495} (\bibinfo {year} {2002})},\ \Eprint
  {http://arxiv.org/abs/hep-lat/0110139} {arXiv:hep-lat/0110139 [hep-lat]}
  \BibitemShut {NoStop}%
\bibitem [{\citenamefont {de~Forcrand}\ and\ \citenamefont {von
  Smekal}(2002)}]{deForcrand:2001nd}%
  \BibitemOpen
  \bibfield  {author} {\bibinfo {author} {\bibfnamefont {P.}~\bibnamefont
  {de~Forcrand}}\ and\ \bibinfo {author} {\bibfnamefont {L.}~\bibnamefont {von
  Smekal}},\ }\href {\doibase 10.1103/PhysRevD.66.011504} {\bibfield  {journal}
  {\bibinfo  {journal} {Phys.Rev.}\ }\textbf {\bibinfo {volume} {D66}},\
  \bibinfo {pages} {011504} (\bibinfo {year} {2002})},\ \Eprint
  {http://arxiv.org/abs/hep-lat/0107018} {arXiv:hep-lat/0107018 [hep-lat]}
  \BibitemShut {NoStop}%
\bibitem [{\citenamefont {Barresi}\ \emph {et~al.}(2003)\citenamefont
  {Barresi}, \citenamefont {Burgio},\ and\ \citenamefont
  {Muller-Preussker}}]{Barresi:2002un}%
  \BibitemOpen
  \bibfield  {author} {\bibinfo {author} {\bibfnamefont {A.}~\bibnamefont
  {Barresi}}, \bibinfo {author} {\bibfnamefont {G.}~\bibnamefont {Burgio}}, \
  and\ \bibinfo {author} {\bibfnamefont {M.}~\bibnamefont {Muller-Preussker}},\
  }\href {\doibase 10.1016/S0920-5632(03)01623-2} {\bibfield  {journal}
  {\bibinfo  {journal} {Nucl.Phys.Proc.Suppl.}\ }\textbf {\bibinfo {volume}
  {119}},\ \bibinfo {pages} {571} (\bibinfo {year} {2003})},\ \Eprint
  {http://arxiv.org/abs/hep-lat/0209011} {arXiv:hep-lat/0209011 [hep-lat]}
  \BibitemShut {NoStop}%
\bibitem [{\citenamefont {Reinhardt}(2003)}]{Reinhardt:2002mb}%
  \BibitemOpen
  \bibfield  {author} {\bibinfo {author} {\bibfnamefont {H.}~\bibnamefont
  {Reinhardt}},\ }\href {\doibase 10.1016/S0370-2693(03)00199-0} {\bibfield
  {journal} {\bibinfo  {journal} {Phys.Lett.}\ }\textbf {\bibinfo {volume}
  {B557}},\ \bibinfo {pages} {317} (\bibinfo {year} {2003})},\ \Eprint
  {http://arxiv.org/abs/hep-th/0212264} {arXiv:hep-th/0212264 [hep-th]}
  \BibitemShut {NoStop}%
\bibitem [{\citenamefont {Barresi}\ \emph
  {et~al.}(2004{\natexlab{a}})\citenamefont {Barresi}, \citenamefont {Burgio},\
  and\ \citenamefont {Muller-Preussker}}]{Barresi:2003jq}%
  \BibitemOpen
  \bibfield  {author} {\bibinfo {author} {\bibfnamefont {A.}~\bibnamefont
  {Barresi}}, \bibinfo {author} {\bibfnamefont {G.}~\bibnamefont {Burgio}}, \
  and\ \bibinfo {author} {\bibfnamefont {M.}~\bibnamefont {Muller-Preussker}},\
  }\href {\doibase 10.1103/PhysRevD.69.094503} {\bibfield  {journal} {\bibinfo
  {journal} {Phys.Rev.}\ }\textbf {\bibinfo {volume} {D69}},\ \bibinfo {pages}
  {094503} (\bibinfo {year} {2004}{\natexlab{a}})},\ \Eprint
  {http://arxiv.org/abs/hep-lat/0309010} {arXiv:hep-lat/0309010 [hep-lat]}
  \BibitemShut {NoStop}%
\bibitem [{\citenamefont {Barresi}\ \emph
  {et~al.}(2004{\natexlab{b}})\citenamefont {Barresi}, \citenamefont {Burgio},
  \citenamefont {D'Elia},\ and\ \citenamefont
  {Muller-Preussker}}]{Barresi:2004qa}%
  \BibitemOpen
  \bibfield  {author} {\bibinfo {author} {\bibfnamefont {A.}~\bibnamefont
  {Barresi}}, \bibinfo {author} {\bibfnamefont {G.}~\bibnamefont {Burgio}},
  \bibinfo {author} {\bibfnamefont {M.}~\bibnamefont {D'Elia}}, \ and\ \bibinfo
  {author} {\bibfnamefont {M.}~\bibnamefont {Muller-Preussker}},\ }\href
  {\doibase 10.1016/j.physletb.2004.08.051} {\bibfield  {journal} {\bibinfo
  {journal} {Phys.Lett.}\ }\textbf {\bibinfo {volume} {B599}},\ \bibinfo
  {pages} {278} (\bibinfo {year} {2004}{\natexlab{b}})},\ \Eprint
  {http://arxiv.org/abs/hep-lat/0405004} {arXiv:hep-lat/0405004 [hep-lat]}
  \BibitemShut {NoStop}%
\bibitem [{\citenamefont {Barresi}\ and\ \citenamefont
  {Burgio}(2007)}]{Barresi:2006gq}%
  \BibitemOpen
  \bibfield  {author} {\bibinfo {author} {\bibfnamefont {A.}~\bibnamefont
  {Barresi}}\ and\ \bibinfo {author} {\bibfnamefont {G.}~\bibnamefont
  {Burgio}},\ }\href {\doibase 10.1140/epjc/s10052-006-0172-8} {\bibfield
  {journal} {\bibinfo  {journal} {Eur.Phys.J.}\ }\textbf {\bibinfo {volume}
  {C49}},\ \bibinfo {pages} {973} (\bibinfo {year} {2007})},\ \Eprint
  {http://arxiv.org/abs/hep-lat/0608008} {arXiv:hep-lat/0608008 [hep-lat]}
  \BibitemShut {NoStop}%
\bibitem [{\citenamefont {Burgio}\ \emph {et~al.}(2006)\citenamefont {Burgio},
  \citenamefont {Fuhrmann}, \citenamefont {Kerler},\ and\ \citenamefont
  {Muller-Preussker}}]{Burgio:2006dc}%
  \BibitemOpen
  \bibfield  {author} {\bibinfo {author} {\bibfnamefont {G.}~\bibnamefont
  {Burgio}}, \bibinfo {author} {\bibfnamefont {M.}~\bibnamefont {Fuhrmann}},
  \bibinfo {author} {\bibfnamefont {W.}~\bibnamefont {Kerler}}, \ and\ \bibinfo
  {author} {\bibfnamefont {M.}~\bibnamefont {Muller-Preussker}},\ }\href
  {\doibase 10.1103/PhysRevD.74.071502} {\bibfield  {journal} {\bibinfo
  {journal} {Phys.Rev.}\ }\textbf {\bibinfo {volume} {D74}},\ \bibinfo {pages}
  {071502} (\bibinfo {year} {2006})},\ \Eprint
  {http://arxiv.org/abs/hep-th/0608075} {arXiv:hep-th/0608075 [hep-th]}
  \BibitemShut {NoStop}%
\bibitem [{\citenamefont {Burgio}\ \emph {et~al.}(2007)\citenamefont {Burgio},
  \citenamefont {Fuhrmann}, \citenamefont {Kerler},\ and\ \citenamefont
  {Muller-Preussker}}]{Burgio:2006xj}%
  \BibitemOpen
  \bibfield  {author} {\bibinfo {author} {\bibfnamefont {G.}~\bibnamefont
  {Burgio}}, \bibinfo {author} {\bibfnamefont {M.}~\bibnamefont {Fuhrmann}},
  \bibinfo {author} {\bibfnamefont {W.}~\bibnamefont {Kerler}}, \ and\ \bibinfo
  {author} {\bibfnamefont {M.}~\bibnamefont {Muller-Preussker}},\ }\href
  {\doibase 10.1103/PhysRevD.75.014504} {\bibfield  {journal} {\bibinfo
  {journal} {Phys.Rev.}\ }\textbf {\bibinfo {volume} {D75}},\ \bibinfo {pages}
  {014504} (\bibinfo {year} {2007})},\ \Eprint
  {http://arxiv.org/abs/hep-lat/0610097} {arXiv:hep-lat/0610097 [hep-lat]}
  \BibitemShut {NoStop}%
\bibitem [{\citenamefont {Burgio}\ and\ \citenamefont
  {Reinhardt}(2015)}]{Burgio:2014yna}%
  \BibitemOpen
  \bibfield  {author} {\bibinfo {author} {\bibfnamefont {G.}~\bibnamefont
  {Burgio}}\ and\ \bibinfo {author} {\bibfnamefont {H.}~\bibnamefont
  {Reinhardt}},\ }\href {\doibase 10.1103/PhysRevD.91.025021} {\bibfield
  {journal} {\bibinfo  {journal} {Phys.Rev.}\ }\textbf {\bibinfo {volume}
  {D91}},\ \bibinfo {pages} {025021} (\bibinfo {year} {2015})},\ \Eprint
  {http://arxiv.org/abs/1412.1762} {arXiv:1412.1762 [hep-lat]} \BibitemShut
  {NoStop}%
\bibitem [{\citenamefont {Klassen}(1998)}]{Klassen:1998ua}%
  \BibitemOpen
  \bibfield  {author} {\bibinfo {author} {\bibfnamefont {T.~R.}\ \bibnamefont
  {Klassen}},\ }\href {\doibase 10.1016/S0550-3213(98)00510-0} {\bibfield
  {journal} {\bibinfo  {journal} {Nucl.Phys.}\ }\textbf {\bibinfo {volume}
  {B533}},\ \bibinfo {pages} {557} (\bibinfo {year} {1998})},\ \Eprint
  {http://arxiv.org/abs/hep-lat/9803010} {arXiv:hep-lat/9803010 [hep-lat]}
  \BibitemShut {NoStop}%
\bibitem [{\citenamefont {Schr\"ock}\ and\ \citenamefont
  {Vogt}(2013)}]{Schrock:2012fj}%
  \BibitemOpen
  \bibfield  {author} {\bibinfo {author} {\bibfnamefont {M.}~\bibnamefont
  {Schr\"ock}}\ and\ \bibinfo {author} {\bibfnamefont {H.}~\bibnamefont
  {Vogt}},\ }\href {\doibase 10.1016/j.cpc.2013.03.021} {\bibfield  {journal}
  {\bibinfo  {journal} {Comput.Phys.Commun.}\ }\textbf {\bibinfo {volume}
  {184}},\ \bibinfo {pages} {1907} (\bibinfo {year} {2013})},\ \Eprint
  {http://arxiv.org/abs/1212.5221} {arXiv:1212.5221 [hep-lat]} \BibitemShut
  {NoStop}%
\bibitem [{\citenamefont {Bali}\ \emph {et~al.}(1996)\citenamefont {Bali},
  \citenamefont {Bornyakov}, \citenamefont {Muller-Preussker},\ and\
  \citenamefont {Schilling}}]{Bali:1996dm}%
  \BibitemOpen
  \bibfield  {author} {\bibinfo {author} {\bibfnamefont {G.}~\bibnamefont
  {Bali}}, \bibinfo {author} {\bibfnamefont {V.}~\bibnamefont {Bornyakov}},
  \bibinfo {author} {\bibfnamefont {M.}~\bibnamefont {Muller-Preussker}}, \
  and\ \bibinfo {author} {\bibfnamefont {K.}~\bibnamefont {Schilling}},\ }\href
  {\doibase 10.1103/PhysRevD.54.2863} {\bibfield  {journal} {\bibinfo
  {journal} {Phys.Rev.}\ }\textbf {\bibinfo {volume} {D54}},\ \bibinfo {pages}
  {2863} (\bibinfo {year} {1996})},\ \Eprint
  {http://arxiv.org/abs/hep-lat/9603012} {arXiv:hep-lat/9603012 [hep-lat]}
  \BibitemShut {NoStop}%
\bibitem [{\citenamefont {Bornyakov}\ \emph {et~al.}(2001)\citenamefont
  {Bornyakov}, \citenamefont {Komarov},\ and\ \citenamefont
  {Polikarpov}}]{Bornyakov:2000ig}%
  \BibitemOpen
  \bibfield  {author} {\bibinfo {author} {\bibfnamefont {V.}~\bibnamefont
  {Bornyakov}}, \bibinfo {author} {\bibfnamefont {D.}~\bibnamefont {Komarov}},
  \ and\ \bibinfo {author} {\bibfnamefont {M.}~\bibnamefont {Polikarpov}},\
  }\href {\doibase 10.1016/S0370-2693(00)01309-5} {\bibfield  {journal}
  {\bibinfo  {journal} {Phys.Lett.}\ }\textbf {\bibinfo {volume} {B497}},\
  \bibinfo {pages} {151} (\bibinfo {year} {2001})},\ \Eprint
  {http://arxiv.org/abs/hep-lat/0009035} {arXiv:hep-lat/0009035 [hep-lat]}
  \BibitemShut {NoStop}%
\bibitem [{\citenamefont {Giusti}\ \emph {et~al.}(2001)\citenamefont {Giusti},
  \citenamefont {Paciello}, \citenamefont {Parrinello}, \citenamefont
  {Petrarca},\ and\ \citenamefont {Taglienti}}]{Giusti:2001xf}%
  \BibitemOpen
  \bibfield  {author} {\bibinfo {author} {\bibfnamefont {L.}~\bibnamefont
  {Giusti}}, \bibinfo {author} {\bibfnamefont {M.}~\bibnamefont {Paciello}},
  \bibinfo {author} {\bibfnamefont {C.}~\bibnamefont {Parrinello}}, \bibinfo
  {author} {\bibfnamefont {S.}~\bibnamefont {Petrarca}}, \ and\ \bibinfo
  {author} {\bibfnamefont {B.}~\bibnamefont {Taglienti}},\ }\href {\doibase
  10.1142/S0217751X01004281} {\bibfield  {journal} {\bibinfo  {journal}
  {Int.J.Mod.Phys.}\ }\textbf {\bibinfo {volume} {A16}},\ \bibinfo {pages}
  {3487} (\bibinfo {year} {2001})},\ \Eprint
  {http://arxiv.org/abs/hep-lat/0104012} {arXiv:hep-lat/0104012 [hep-lat]}
  \BibitemShut {NoStop}%
\bibitem [{\citenamefont {Faber}\ \emph
  {et~al.}(2001{\natexlab{a}})\citenamefont {Faber}, \citenamefont
  {Greensite},\ and\ \citenamefont {Olejnik}}]{Faber:2001hq}%
  \BibitemOpen
  \bibfield  {author} {\bibinfo {author} {\bibfnamefont {M.}~\bibnamefont
  {Faber}}, \bibinfo {author} {\bibfnamefont {J.}~\bibnamefont {Greensite}}, \
  and\ \bibinfo {author} {\bibfnamefont {S.}~\bibnamefont {Olejnik}},\ }\href
  {\doibase 10.1103/PhysRevD.64.034511} {\bibfield  {journal} {\bibinfo
  {journal} {Phys.Rev.}\ }\textbf {\bibinfo {volume} {D64}},\ \bibinfo {pages}
  {034511} (\bibinfo {year} {2001}{\natexlab{a}})},\ \Eprint
  {http://arxiv.org/abs/hep-lat/0103030} {arXiv:hep-lat/0103030 [hep-lat]}
  \BibitemShut {NoStop}%
\bibitem [{\citenamefont {Faber}\ \emph
  {et~al.}(2001{\natexlab{b}})\citenamefont {Faber}, \citenamefont
  {Greensite},\ and\ \citenamefont {Olejnik}}]{Faber:2001zs}%
  \BibitemOpen
  \bibfield  {author} {\bibinfo {author} {\bibfnamefont {M.}~\bibnamefont
  {Faber}}, \bibinfo {author} {\bibfnamefont {J.}~\bibnamefont {Greensite}}, \
  and\ \bibinfo {author} {\bibfnamefont {S.}~\bibnamefont {Olejnik}},\ }\href
  {\doibase 10.1088/1126-6708/2001/11/053} {\bibfield  {journal} {\bibinfo
  {journal} {JHEP}\ }\textbf {\bibinfo {volume} {0111}},\ \bibinfo {pages}
  {053} (\bibinfo {year} {2001}{\natexlab{b}})},\ \Eprint
  {http://arxiv.org/abs/hep-lat/0106017} {arXiv:hep-lat/0106017 [hep-lat]}
  \BibitemShut {NoStop}%
\bibitem [{\citenamefont {Boyko}\ \emph {et~al.}(2006)\citenamefont {Boyko},
  \citenamefont {Bornyakov}, \citenamefont {Ilgenfritz}, \citenamefont
  {Kovalenko}, \citenamefont {Martemyanov} \emph {et~al.}}]{Boyko:2006ic}%
  \BibitemOpen
  \bibfield  {author} {\bibinfo {author} {\bibfnamefont {P.~Y.}\ \bibnamefont
  {Boyko}}, \bibinfo {author} {\bibfnamefont {V.}~\bibnamefont {Bornyakov}},
  \bibinfo {author} {\bibfnamefont {E.-M.}\ \bibnamefont {Ilgenfritz}},
  \bibinfo {author} {\bibfnamefont {A.}~\bibnamefont {Kovalenko}}, \bibinfo
  {author} {\bibfnamefont {B.}~\bibnamefont {Martemyanov}},  \emph {et~al.},\
  }\href {\doibase 10.1016/j.nuclphysb.2006.08.025} {\bibfield  {journal}
  {\bibinfo  {journal} {Nucl.Phys.}\ }\textbf {\bibinfo {volume} {B756}},\
  \bibinfo {pages} {71} (\bibinfo {year} {2006})},\ \Eprint
  {http://arxiv.org/abs/hep-lat/0607003} {arXiv:hep-lat/0607003 [hep-lat]}
  \BibitemShut {NoStop}%
\bibitem [{\citenamefont {O'Cais}\ \emph {et~al.}(2010)\citenamefont {O'Cais},
  \citenamefont {Kamleh}, \citenamefont {Langfeld}, \citenamefont {Lasscock},
  \citenamefont {Leinweber} \emph {et~al.}}]{Cais:2008za}%
  \BibitemOpen
  \bibfield  {author} {\bibinfo {author} {\bibfnamefont {A.}~\bibnamefont
  {O'Cais}}, \bibinfo {author} {\bibfnamefont {W.}~\bibnamefont {Kamleh}},
  \bibinfo {author} {\bibfnamefont {K.}~\bibnamefont {Langfeld}}, \bibinfo
  {author} {\bibfnamefont {B.}~\bibnamefont {Lasscock}}, \bibinfo {author}
  {\bibfnamefont {D.}~\bibnamefont {Leinweber}},  \emph {et~al.},\ }\href
  {\doibase 10.1103/PhysRevD.82.114512} {\bibfield  {journal} {\bibinfo
  {journal} {Phys.Rev.}\ }\textbf {\bibinfo {volume} {D82}},\ \bibinfo {pages}
  {114512} (\bibinfo {year} {2010})},\ \Eprint {http://arxiv.org/abs/0807.0264}
  {arXiv:0807.0264 [hep-lat]} \BibitemShut {NoStop}%
\bibitem [{\citenamefont {Bogolubsky}\ \emph {et~al.}(2006)\citenamefont
  {Bogolubsky}, \citenamefont {Burgio}, \citenamefont {Muller-Preus\-sker},\
  and\ \citenamefont {Mitrjushkin}}]{Bogolubsky:2005wf}%
  \BibitemOpen
  \bibfield  {author} {\bibinfo {author} {\bibfnamefont {I.}~\bibnamefont
  {Bogolubsky}}, \bibinfo {author} {\bibfnamefont {G.}~\bibnamefont {Burgio}},
  \bibinfo {author} {\bibfnamefont {M.}~\bibnamefont {Muller-Preus\-sker}}, \
  and\ \bibinfo {author} {\bibfnamefont {V.}~\bibnamefont {Mitrjushkin}},\
  }\href {\doibase 10.1103/PhysRevD.74.034503} {\bibfield  {journal} {\bibinfo
  {journal} {Phys.Rev.}\ }\textbf {\bibinfo {volume} {D74}},\ \bibinfo {pages}
  {034503} (\bibinfo {year} {2006})},\ \Eprint
  {http://arxiv.org/abs/hep-lat/0511056} {arXiv:hep-lat/0511056 [hep-lat]}
  \BibitemShut {NoStop}%
\bibitem [{\citenamefont {Bogolubsky}\ \emph {et~al.}(2008)\citenamefont
  {Bogolubsky}, \citenamefont {Bornyakov}, \citenamefont {Burgio},
  \citenamefont {Ilgen\-fritz}, \citenamefont {Muller-Preu\-ssker} \emph
  {et~al.}}]{Bogolubsky:2007bw}%
  \BibitemOpen
  \bibfield  {author} {\bibinfo {author} {\bibfnamefont {I.}~\bibnamefont
  {Bogolubsky}}, \bibinfo {author} {\bibfnamefont {V.}~\bibnamefont
  {Bornyakov}}, \bibinfo {author} {\bibfnamefont {G.}~\bibnamefont {Burgio}},
  \bibinfo {author} {\bibfnamefont {E.}~\bibnamefont {Ilgen\-fritz}}, \bibinfo
  {author} {\bibfnamefont {M.}~\bibnamefont {Muller-Preu\-ssker}},  \emph
  {et~al.},\ }\href {\doibase 10.1103/PhysRevD.77.014504,
  10.1103/PhysRevD.77.039902, 10.1103/PhysRevD.77.014504,
  10.1103/PhysRevD.77.039902} {\bibfield  {journal} {\bibinfo  {journal}
  {Phys.Rev.}\ }\textbf {\bibinfo {volume} {D77}},\ \bibinfo {pages} {014504}
  (\bibinfo {year} {2008})},\ \Eprint {http://arxiv.org/abs/0707.3611}
  {arXiv:0707.3611 [hep-lat]} \BibitemShut {NoStop}%
\bibitem [{\citenamefont {Marinari}\ \emph {et~al.}(1993)\citenamefont
  {Marinari}, \citenamefont {Paciello}, \citenamefont {Parisi},\ and\
  \citenamefont {Taglienti}}]{Marinari:1992kh}%
  \BibitemOpen
  \bibfield  {author} {\bibinfo {author} {\bibfnamefont {E.}~\bibnamefont
  {Marinari}}, \bibinfo {author} {\bibfnamefont {M.~L.}\ \bibnamefont
  {Paciello}}, \bibinfo {author} {\bibfnamefont {G.}~\bibnamefont {Parisi}}, \
  and\ \bibinfo {author} {\bibfnamefont {B.}~\bibnamefont {Taglienti}},\ }\href
  {\doibase 10.1016/0370-2693(93)91840-J} {\bibfield  {journal} {\bibinfo
  {journal} {Phys.Lett.}\ }\textbf {\bibinfo {volume} {B298}},\ \bibinfo
  {pages} {400} (\bibinfo {year} {1993})},\ \Eprint
  {http://arxiv.org/abs/hep-lat/9210021} {arXiv:hep-lat/9210021 [hep-lat]}
  \BibitemShut {NoStop}%
\bibitem [{\citenamefont {Creutz}(1980)}]{Creutz:1980wj}%
  \BibitemOpen
  \bibfield  {author} {\bibinfo {author} {\bibfnamefont {M.}~\bibnamefont
  {Creutz}},\ }\href {\doibase 10.1103/PhysRevLett.45.313} {\bibfield
  {journal} {\bibinfo  {journal} {Phys.Rev.Lett.}\ }\textbf {\bibinfo {volume}
  {45}},\ \bibinfo {pages} {313} (\bibinfo {year} {1980})}\BibitemShut
  {NoStop}%
\bibitem [{\citenamefont {Gonzalez-Arroyo}\ and\ \citenamefont
  {Okawa}(2013)}]{GonzalezArroyo:2012fx}%
  \BibitemOpen
  \bibfield  {author} {\bibinfo {author} {\bibfnamefont {A.}~\bibnamefont
  {Gonzalez-Arroyo}}\ and\ \bibinfo {author} {\bibfnamefont {M.}~\bibnamefont
  {Okawa}},\ }\href {\doibase 10.1016/j.physletb.2012.12.027} {\bibfield
  {journal} {\bibinfo  {journal} {Phys.Lett.}\ }\textbf {\bibinfo {volume}
  {B718}},\ \bibinfo {pages} {1524} (\bibinfo {year} {2013})},\ \Eprint
  {http://arxiv.org/abs/1206.0049} {arXiv:1206.0049 [hep-th]} \BibitemShut
  {NoStop}%
\bibitem [{\citenamefont {Albanese}\ \emph {et~al.}(1987)\citenamefont
  {Albanese} \emph {et~al.}}]{Albanese:1987ds}%
  \BibitemOpen
  \bibfield  {author} {\bibinfo {author} {\bibfnamefont {M.}~\bibnamefont
  {Albanese}} \emph {et~al.} (\bibinfo {collaboration} {APE Collaboration}),\
  }\href {\doibase 10.1016/0370-2693(87)91160-9} {\bibfield  {journal}
  {\bibinfo  {journal} {Phys.Lett.}\ }\textbf {\bibinfo {volume} {B192}},\
  \bibinfo {pages} {163} (\bibinfo {year} {1987})}\BibitemShut {NoStop}%
\bibitem [{\citenamefont {Bloch}\ \emph {et~al.}(2004)\citenamefont {Bloch},
  \citenamefont {Cucchieri}, \citenamefont {Langfeld},\ and\ \citenamefont
  {Mendes}}]{Bloch:2003sk}%
  \BibitemOpen
  \bibfield  {author} {\bibinfo {author} {\bibfnamefont {J.~C.~R.}\
  \bibnamefont {Bloch}}, \bibinfo {author} {\bibfnamefont {A.}~\bibnamefont
  {Cucchieri}}, \bibinfo {author} {\bibfnamefont {K.}~\bibnamefont {Langfeld}},
  \ and\ \bibinfo {author} {\bibfnamefont {T.}~\bibnamefont {Mendes}},\ }\href
  {\doibase 10.1016/j.nuclphysb.2004.03.021} {\bibfield  {journal} {\bibinfo
  {journal} {Nucl. Phys.}\ }\textbf {\bibinfo {volume} {B687}},\ \bibinfo
  {pages} {76} (\bibinfo {year} {2004})},\ \Eprint
  {http://arxiv.org/abs/hep-lat/0312036} {arXiv:hep-lat/0312036} \BibitemShut
  {NoStop}%
\end{thebibliography}%

\end{document}